\documentclass[aps,pre,groupedaddress,twocolumn, showpacs]{revtex4}
\usepackage{graphicx}
\usepackage{graphics}
\usepackage{psfrag}
\usepackage{color}
\usepackage{subfigure}
\usepackage{amsmath}
\usepackage{natbib}
\usepackage{epic}
\usepackage{latexsym}
\usepackage{amssymb}
\usepackage{epic}
\newcommand{\etal}{{\it et al.}~}
\newcommand{\btheta}{\boldsymbol{\theta}}
\begin{document}
\title[Bayesian inference for the Neolithisation of Europe]{Bayesian inference
for a wavefront model of the Neolithisation of Europe}

\author{Andrew W.\ Baggaley}
\email{a.w.baggaley@ncl.ac.uk}
\author{Graeme R.\ Sarson}
\author{Anvar Shukurov}
\author{Richard J.\ Boys}
\author{Andrew Golightly}

\affiliation{School of Mathematics and Statistics,
Newcastle University, Newcastle upon Tyne, NE1 7RU, England, UK}

\date{\today}
\begin{abstract}
  We consider a wavefront model for the spread of Neolithic culture
  across Europe, and use Bayesian inference techniques to provide
  estimates for the parameters within this model, as constrained by
  radiocarbon data from Southern and Western Europe.  Our wavefront
  model allows for both an isotropic background spread (incorporating
  the effects of local geography), and a localized anisotropic spread
  associated with major waterways.  We introduce an innovative
  numerical scheme to track the wavefront, and use Gaussian process emulators 
  to further increase the efficiency of our model, thereby making 
  Markov chain Monte Carlo methods practical.  
  We allow for uncertainty in the fit of our model, and discuss the 
  inferred distribution of the parameter specifying this uncertainty, 
  along with the distributions of the parameters of our wavefront model.
  We subsequently use predictive distributions, taking account of parameter
  uncertainty, to identify radiocarbon sites which do not agree well
  with our model.  These sites may warrant further archaeological
  study, or motivate refinements to the model.
\end{abstract}

\maketitle

\section{Introduction}
\label{sect:intro}

The transition from hunter-gathering to early farming ---
signifying the start of the Neolithic era 
in traditional archaeological terminology ---
was one of the most important steps made by humanity 
in developing the complex modern societies that exist today. 
The mechanism of the spread throughout Europe of Neolithic farming techniques, 
which developed in the Near East around 12,000 years ago,
remains an important and fascinating question.
The relative importance of `cultural' versus `demic' components of the
spread --- i.e.\ the transmission of farming techniques 
versus the physical migration of farmers ---
has long been debated in the archaeological literature.
Intriguingly, advances in genetics mean that quantitative assessments 
of these issues are now becoming possible,
with many recent studies suggesting at least some migration of early farmers
(e.g.\ \cite{Burger:2011,Skoglund:2012}).

Recently there have been a number of studies using population dynamics
models to describe the spread of Neolithic farmers.  Whilst some
recent work has focused on stochastic methods \cite{Fedotov:2008},
most studies have built upon the pioneering work of Ammerman \&
Cavalli-Sforza \cite{Ammerman:1971}, and sought the solution of a
deterministic partial differential equation
\cite{Ackland:2007,Fort:1999,Fort:2010}.  The success of this approach
can be seen in a number of studies.  For example, when using values
for the model inputs determined theoretically or from the
archaeological and anthropological literature,
\cite{Davison:2006,Davison:2009} found a reasonable agreement between
the output of the numerical model and the large-scale features of the
observed first arrival times at Neolithic sites, based on the
radiocarbon dates of objects found at these sites. Unfortunately there
may be many other parameter sets which provide equally if not better
fitting model outputs to the data.  We seek these parameters sets by
developing a rigorous statistical inference method to fit the model to
the radiocarbon data.

In this paper we adopt a Bayesian approach to inference as this will
also allow us to quantify parameter uncertainty in a rigorous way.
Additionally it allows us to quantify correlations between our model
parameters, and the global uncertainty in the fit of our model to the
data.  The authors are not aware of another study where these
sophisticated statistical techniques have been used, allowing the
determination not just of parameter estimates, but of a plausible
range of parameter values, given by the posterior distribution.

The `wave of advance' is one of the most important concepts in
modeling the spread of the Neolithic, underlying the studies cited
above; since the work of Ammerman \& Cavalli-Sforza
\citep{Ammerman:1971} and Clark \citep{Clark:1965}, it has been widely
accepted that the incipient farming spread from the Levant to Western
Europe in a systematic manner --- an outward propagating `wave' ---
amenable to study by simple deterministic models.  The rate of
propagation of the wave is broadly constant throughout this spread,
and various authors have estimated the speed of the wave front, $U$,
from the radiocarbon data.  In the present work, we try to produce a
statistically reliable estimate of the wave speed $U$ from the
radiocarbon data.  Unlike most if not all earlier studies, however, we
explicitly account for the fact that $U$ is a random variable, due
both to the systematic and random errors inherent in the data, and to
the inherent variability in the underlying population dynamics
processes.  (The true spread is not well-modeled on all scales, and at
all locations, by a wavefront advancing with a continuous speed; local
variations do of course occur.  Such local variations are not
explicitly modeled within simple deterministic wavefront models, which
effectively average out such small-scale variability, and reproduce
the spread well on the larger, continental scales.)

As well as being the natural quantity to describe the spread, 
the wave speed has clear and important implications 
within most mathematical models of the spread.
For example, 
within the Fisher--Kolmogorov--Petrovsky--Piskunov (FKPP) equation
\citep{Fisher:1937,Kolmogorov:1937}, 
most frequently used in models of demic diffusion,
the front speed is directly linked to 
the population mobility (diffusivity) and growth rate.
While the wave front has been most often studied within the context
of the FKPP equation, 
many alternative models of populations dynamics also predict waves of advance,
with speeds dependent upon their various model parameters.
For example, within multiple-population models related to the FKPP model 
but allowing for cultural conversion,
the speed can be affected by the parameters controlling the conversion
\citep{Aoki:1996,Patterson:2010};
similarly, within other models of cultural transmission, 
it can be linked to the intensity and spatial range of contacts 
\citep{Cavalli-Sforza:1981}.
The same is true for many alternative models:
e.g.\ models involving alternative parameterizations 
of growth processes and diffusion \citep{Cohen:1992,Ackland:2007};
and models involving non-Laplacian diffusion 
(e.g.\ L\'evy flights; \cite{Negrete:2003}). 
Even within FKPP-like models with logistic growth, 
`time-delay' factors attempting to model 
the generational effects of population growth more realistically
result in a modified relationship between the front speed and the
basic demographic parameters \citep{Fort:1999,Fort:2010}.
The key quantity which must be inferred from the radiocarbon data 
is the wave speed, 
and so a wavefront-based approach such as that introduced here, 
which gives this quantity prominence,
is a more natural model.

Edmonson \citep{Edmonson:1961} was the first to estimate the
speed of the agropastoral transmission in Europe; 
he gave a value of 1.9\,km/year.
Later, Ammerman \& Cavalli-Sforza \citep{Ammerman:1971} 
gave a value of 1\,km/year,
and most subsequent determinations of the speed of the Neolithic wavefront 
have also been of order $U\simeq1\,$km/year \cite{Ackland:2007,Davison:2006}.
There are notable regional variations, however. 
In particular, the Linearbandkeramik (LBK) culture spread along
the Danube--Rhine corridor at a higher speed, perhaps as large as 5\,km/year
\cite{Zilhao:2001,Dolukhanov:2005}.
The spread of the Impressed Ware ceramics along the Mediterranean coast
has been estimated to be as fast as 10\,km/year \citep{Zilhao:2001},
although lower estimates are also possible \citep{Zilhao:2003}.
In contrast, there is no clear evidence for any significant acceleration 
along the northern and Atlantic coastline of Europe.
In this paper we follow \cite{Davison:2006} in allowing for 
enhanced anisotropic spread along coastlines 
and along the courses of the Danube and Rhine rivers,
in an attempt to model the above phenomena.
We perform Bayesian inference for the magnitudes
of these enhanced effects, 
and for the magnitude of the isotropic background spread;
as a result, we are in a position to assess the extent 
to which these features of the model are genuinely required 
by the radiocarbon data.

Although the posterior distribution of the parameters of interest is
analytically intractable, computationally intensive methods such as
Markov chain Monte Carlo (MCMC) can be used to generate samples from
this distribution. The computational cost of simulating the wave of
advance precludes the direct use of an MCMC scheme.  We therefore
approximate the arrival time of the wavefront at each site using a
Gaussian process emulator \citep{SantnerWN03} as the computational
speed-up makes MCMC methods practicable.

The rest of this paper is organized as follows.
We introduce our wavefront model, 
and compare its output to that obtained from a more traditional approach
involving partial differential equations (PDEs),
in section~\ref{sec:particles}.
In section~\ref{sec:rcarb} we describe the data against which our model 
will be compared.
We outline our statistical model, 
and the Bayesian inference scheme used to estimate our model parameters 
from the data, in section~\ref{sec:MCMC}.
The results of our inference are described in section~\ref{sec:results},
and our conclusions summarized in section~\ref{sec:conc}.
Some technical details about our statistical methods,
including the use of Gaussian process emulators,
are presented as Appendices.


\section{Modelling the propagating front}\label{sec:particles}

The isotropic FKPP equation,
\begin{equation}\label{eq:FKPP}
  \frac{\partial N}{\partial t}=\gamma N\left(1-\frac{N}{K}\right)+ \nabla \cdot (\nu \nabla N),
\end{equation}
describes the evolution of population density $N(\mathbf{x},t)$ 
at position $\mathbf{x}$ and time $t$,
with the growth rate $\gamma$, diffusivity $\nu$ and carrying capacity $K$
as parameters.
The solution to this equation forms a propagating wave front, 
which travels with a speed, 
\begin{equation}\label{eq:front_speed}
  U=2\sqrt{\nu \gamma},
\end{equation}
which is dependent on both the diffusivity 
and the growth rate of the population, 
but importantly is independent of the carrying capacity.
This result can be readily proved in one dimension \cite{Kolmogorov:1937}, 
and also in two dimensions if the curvature of the front is sufficiently small 
(which is to be expected when the distance from the source is large 
compared with the front width)
\cite{Murray:1993}.
In a spatially heterogeneous environment 
(such that $\nu$ and/or $\gamma$ vary with $\mathbf{x}$),
the wavespeed is clearly a function of position, $U(\mathbf{x})$.

Numerical solutions to the FKPP partial differential equation 
are most simply obtained by 
discretizing to a grid of points in space,
using e.g.\ finite difference methods to approximate the spatial derivatives,
and time-stepping the solution forward in time.
Thus the local population density at each point on the grid 
is calculated at each time-step.
If we are focusing on the first arrival time of the Neolithic farmers 
at specific radiocarbon sites, however, 
then all modeling of the population behind the front is unnecessary;
as indeed is the modeling of the equation ahead of the front,
which solution of the FKPP equation also requires.
A reasonable alternative is to model only the propagating front itself, 
which we do using a particle-based approach.
This approach necessarily omits many complex demographic effects 
which may have occurred locally within the real spread;
but that is also true of the FKPP equation most frequently used 
to model the Neolithisation.
And given the limitations of the current data, it would be premature to adopt
more complex demographic models for the spread on the continental scale;
almost all models of the Neolithisation of Europe as a whole have
therefore similarly focused on the propagation of the front.

We select a starting point for our initial population, 
and at a small radius from this point 
we approximate a circle with a small number of `particles' (points); 
these particles define our front.
An alternative approach is to define a regional source, however for simplicity
here we take a localised source.
We keep track of the index of the adjacent particles and so can easily define local tangent and normal vectors,
and in particular the local unit outward normal $\mathbf{\hat{n}}$.
(Here outward means in the direction of the advancing front.)
We perform all simulations in spherical polar coordinates, at a fixed radius, $R$, set to approximate the Earth's surface ($R=6378$ km),
and define the position in terms of the polar and azimuthal angles, 
$\phi$ and $\lambda$, respectively;
i.e.\ $\mathbf{x}=(\phi,\lambda)$.
Numerically, at each time step, we move the particle $i$, 
at position $\mathbf{x}_{i}=(\phi_i,\lambda_i)$
and with velocity $\mathbf{u}_{i}=U(\mathbf{x}_{i}) \hat{\mathbf{n}}_{i}$,
a small amount in both the $\phi$ and $\lambda$ directions
according to
\begin{equation}
\frac{{\rm d}\mathbf{x}_{i}}{{\rm d}t}=\mathbf{u}_{i}.
\end{equation}
Here $\hat{\mathbf{n}}_{i}$ is the local outward normal, and $t$ represents the time in years.
As discussed in section~\ref{sect:intro}, 
there are local
deviations in the rate of spread of the Neolithic, particularly along
traversable waterways. 
We follow \citep{Davison:2006} in allowing an increased
rate of spread along all coastlines, 
and also along the Danube--Rhine river systems. 
We label the enhanced coastal velocity as $\mathbf{V}_{\rm C}$, 
and the river velocity as $\mathbf{V}_{\rm R}$.
In the partial differential equation approach,
these velocities appear as additional advective terms 
in the FKPP equation (Eq.\ \ref{eq:FKPP}), 
which can be identified with anisotropic diffusion \citep{Davison:2006}.
In the wavefront approach,
we can instead simply add this effect to the velocity experienced by particle $i$,
which becomes
\begin{equation}
\label{eq:ui}
\mathbf{u}_{i}=U\mathbf{\hat{n}}_{i} + \mathbf{V}_{i},
\end{equation}
where the total advection from both river and coastal terms
at position $\mathbf{x}_{i}$ is
\begin{equation}
\label{eq:Vi}
\mathbf{V}_{i}=V_{{\rm C}}\textrm{sign}(\hat{\mathbf{n}}_{i} \cdot \widetilde{\mathbf{V}}_{{\rm C},i}) \widetilde{\mathbf{V}}_{{\rm C},i}+V_{\rm R}\textrm{sign}(\hat{\mathbf{n}}_{i} \cdot \widetilde{\mathbf{V}}_{{\rm R},i}) \widetilde{\mathbf{V}}_{{\rm R},i}.
\end{equation}
Here $V_{\rm C/R}$ are the `amplitudes' for the river and coastal advective 
speeds,
and $\widetilde{\mathbf{V}}_{{\rm C/R},i}$ are normalized vectors 
in the direction of the relevant local advection
(normalized to unit magnitude at points on the river or coast).
The sign functions ensure that the sense of the advection 
(e.g.\ upriver or downriver) is that which enhances 
the outward speed of the locally expanding wavefront.
We use MCMC methods below to infer the acceptable range of the
amplitude parameters $V_{\rm C}$ and $V_{\rm R}$,
given the radiocarbon data.

\subsection{Spatial dependence}

It is intuitively sensible that the local altitude 
should have a significant effect on the spread of the Neolithic; 
and, indeed, early farming in Europe does not seem to have been practical 
at altitudes greater than 1\,km above sea level
(e.g., there are no LBK sites above this height 
in the Alpine foreland \cite{Whittle:1996}).
There is also significant evidence \cite{Ammerman:1971,Gkiasta:2003}
that the latitude (acting partly as a proxy for the climate) 
had a significant impact on the productivity of the land, 
and therefore the ability to farm.
Another latitudinal effect noted in the literature 
is an increased competition in the North 
with the pre-existing Mesolithic population \citep{Fort:2010}.
Both of these factors motivate a decreased wavespeed at higher latitudes.

In order to introduce these spatial dependences into the model,
we use arrays of geographical altitude data taken from 
the ETOPO1 1-minute Global Relief database~\citep{geodas}, 
taking a dataset with spatial resolution of 4 arc-minutes.
This forms a 740 by 1100 mesh, with approximate longitudinal boundaries
of $15^\textrm{o}$\,W and $60^\textrm{o}$\,E and latitudinal
boundaries of $25^\textrm{o}$\,N and $75^\textrm{o}$\,N.

\begin{figure}
\begin{center}
    \includegraphics[width=0.45\textwidth]{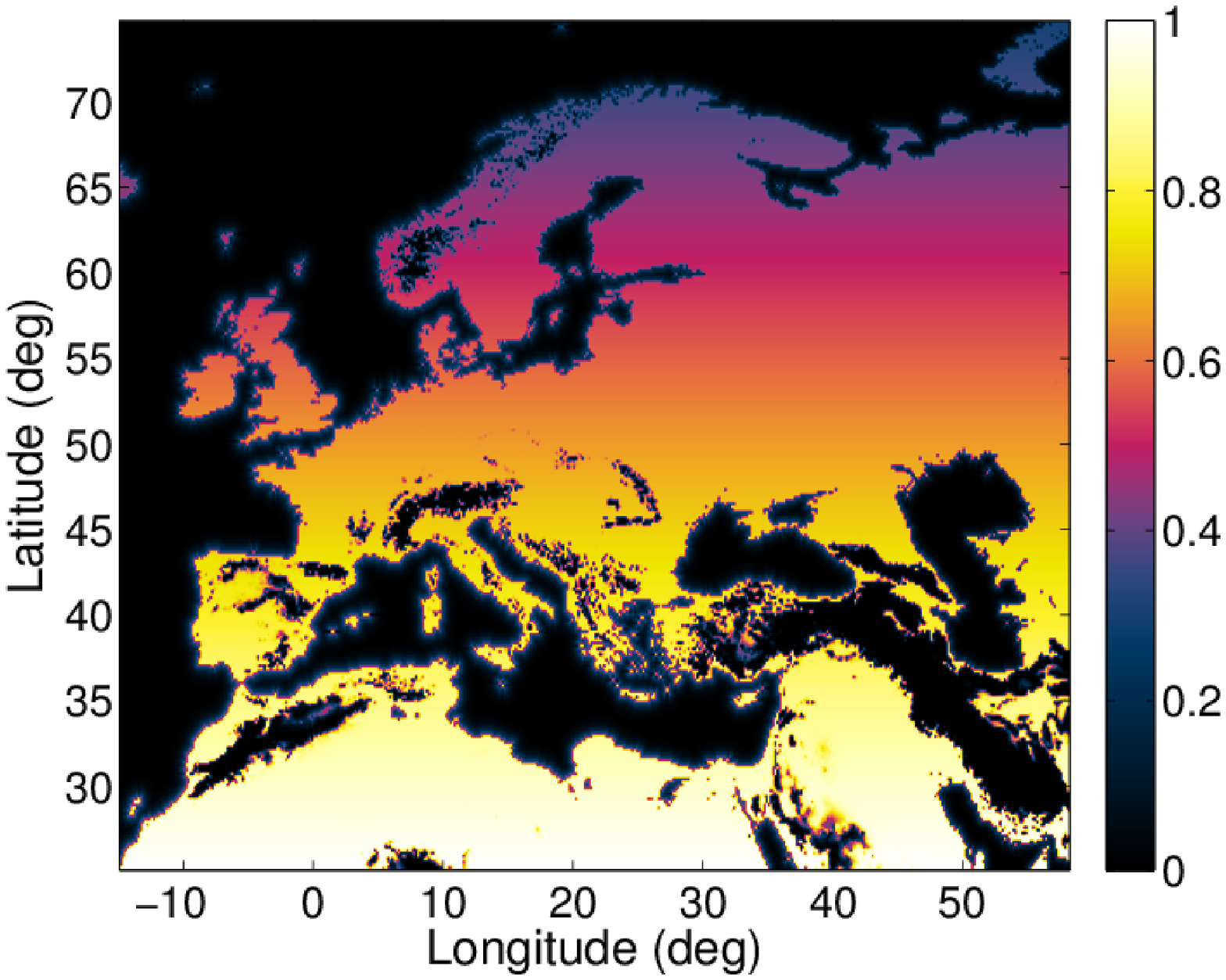}
    \includegraphics[width=0.45\textwidth]{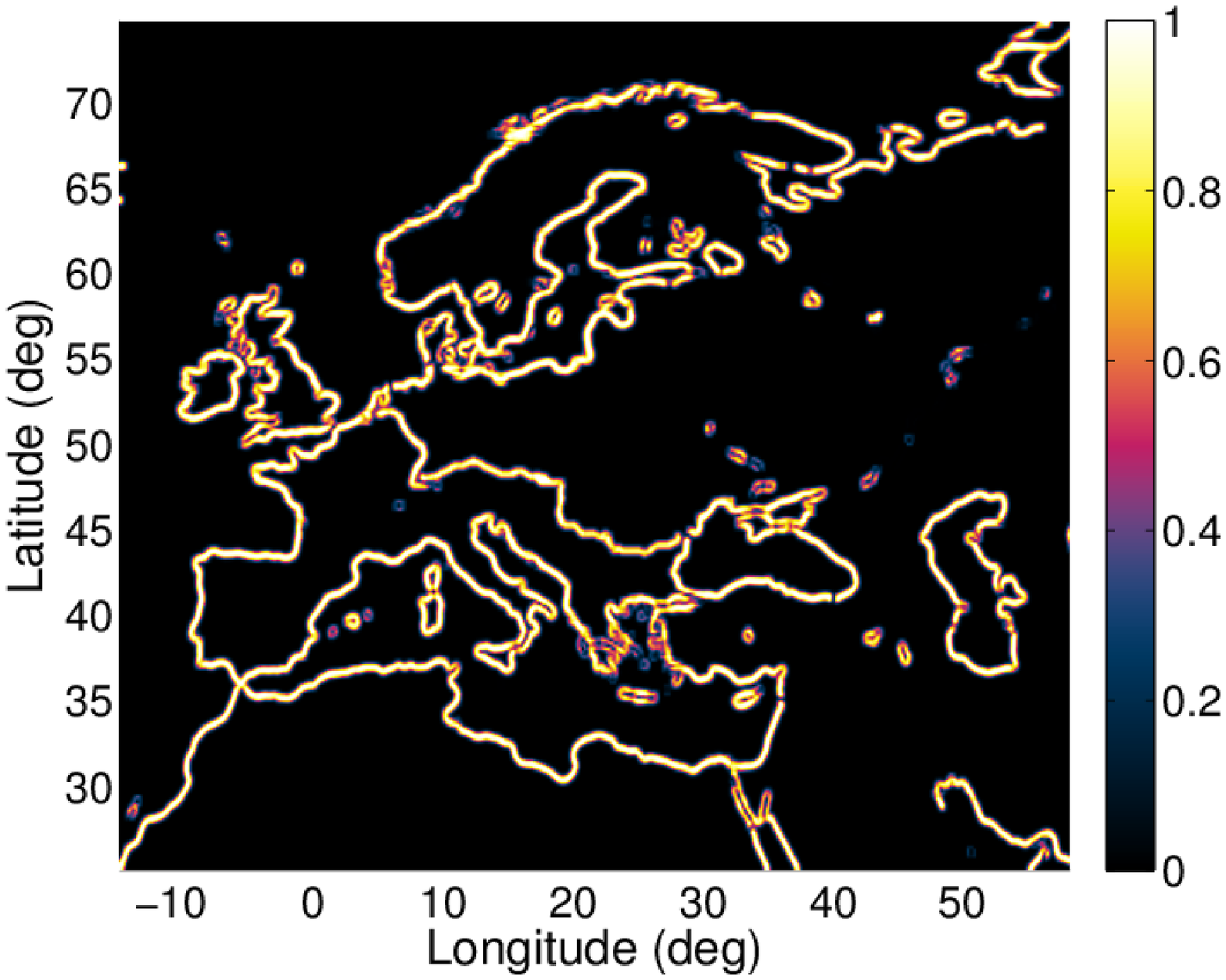}
    \caption{(Color online) (a) Spatial variation of the speed $U(\mathbf{x})$ (from Eq.~\ref{eq:bigU});  (b) magnitude of the distance weighted coastal and river vectors, $|\widetilde{\mathbf{V}}_{i}|$ (as in Eq.~\ref{eq:Vi}).}
\label{fig:nuVR}
\end{center}
\end{figure}

At points on this grid, we calculate the local velocity to reflect the
factors outlined above.  On land, we require the speed to decrease to
zero at altitudes above 1\,km, using a smooth approximation to a step
function.  To allow for limited sea travel we use a different form of
cut-off at low altitude, setting the speed of the wavefront to
decrease exponentially with the distance to the nearest land
($d_\textrm{c}$).  We also include a linear dependence of $U$ on
latitude.  As a result, we calculate the local velocity on our grid as
\begin{equation} \label{eq:bigU}
U = {U_0} \left(\frac{5}{4}-\frac{\phi}{100^\circ}\right)
\begin{cases}
  \frac{1}{2}-\frac{1}{2}\tanh\{10(a-1\,{\rm km})\},& a>0,\\
  \exp(-d_\textrm{c}/10\,{\rm km}), & a<0.
\end{cases}
\end{equation} 
Here $U_0$ is the background amplitude determining the mean rate of
spread; we expect $U_{0}$ to be of order 1\,km/year, but use our MCMC
methods below to infer the acceptable range of this parameter (given
the radiocarbon data) more rigorously, along with the parameters
$V_{\rm C}$ and $V_{\rm R}$ introduced above.  The spatial variations
in $U$ are illustrated in Fig.~\ref{fig:nuVR} (top).

To deal with the advection terms,
the river and coastline vectors used in this study are taken from
\cite{CIA}, which contains vector data of the world's coastlines and
major rivers. 
We take a subset of this data, which contains all the coastlines
within our domain,
and the river vectors corresponding to the Danube and Rhine.
To obtain values of $\mathbf{V}_{\rm C/R}$ at each point on our mesh,
a distance weighted contribution is taken from each of the irregularly spaced vector data segments which define the river in \cite{CIA}.
Specifically, the contribution from each segment
is weighted by $\exp(-d_{\rm vec}/15\,{\rm km})$, where $d_{\rm vec}$ is the
distance between the grid point and the river/coastal vector (in km).
The magnitudes of river/coastal vectors are shown in Fig.~\ref{fig:nuVR} (bottom).
(The small magnitude features visible in some inland non-river locations 
in this plot are associated with small lakes, 
which are included in the coastline database.
These features have no significant effect on our model.) 

It is perhaps worth commenting on our use of present-day information 
about coastline and river locations, since these locatinoteons have 
changed over the timescale of the spread we are modelling;
most obviously, coastlines have changed as a result of sea-level changes
arising principally from post-glacial isostatic adjustment.
During earlier work based on the PDE approach, we investigated 
the effect of such changes, using a sea-level model supplied 
by geophysicists from Durham University
(e.g.\ \cite{Milne:2004,Milne:2006}; Glenn Milne, personal communication).
The effect on our model was negligible, 
since the most significant changes in sea-level were in the North,
and had largely occurred by the time that the Neolithic wave reaches 
the northern coasts
(so that potentially important land bridges had already disappeared)
\citep{DavisonPhD}.
We have not tried to account for changes in the courses of rivers,
but we do not expect that such relatively local changes would have a
significant effect on the large-scale spread on which we focus.

In our wavefront model, to calculate the local speed~$U$ appropriate
at the precise position of particle $i$ (as is needed for
Eq.~\ref{eq:ui}), we use bilinear interpolation from the values at the
four closest mesh-points.  Similarly, the local advective vectors at
the precise position of particle~$i$ are also obtained using bilinear
interpolation from the closest mesh-points; it is these local vectors
which appear in Eq.~(\ref{eq:Vi}).


\subsection{Wavefront algorithms}

\begin{figure*}
\begin{center}
    \includegraphics[width=0.3\textwidth]{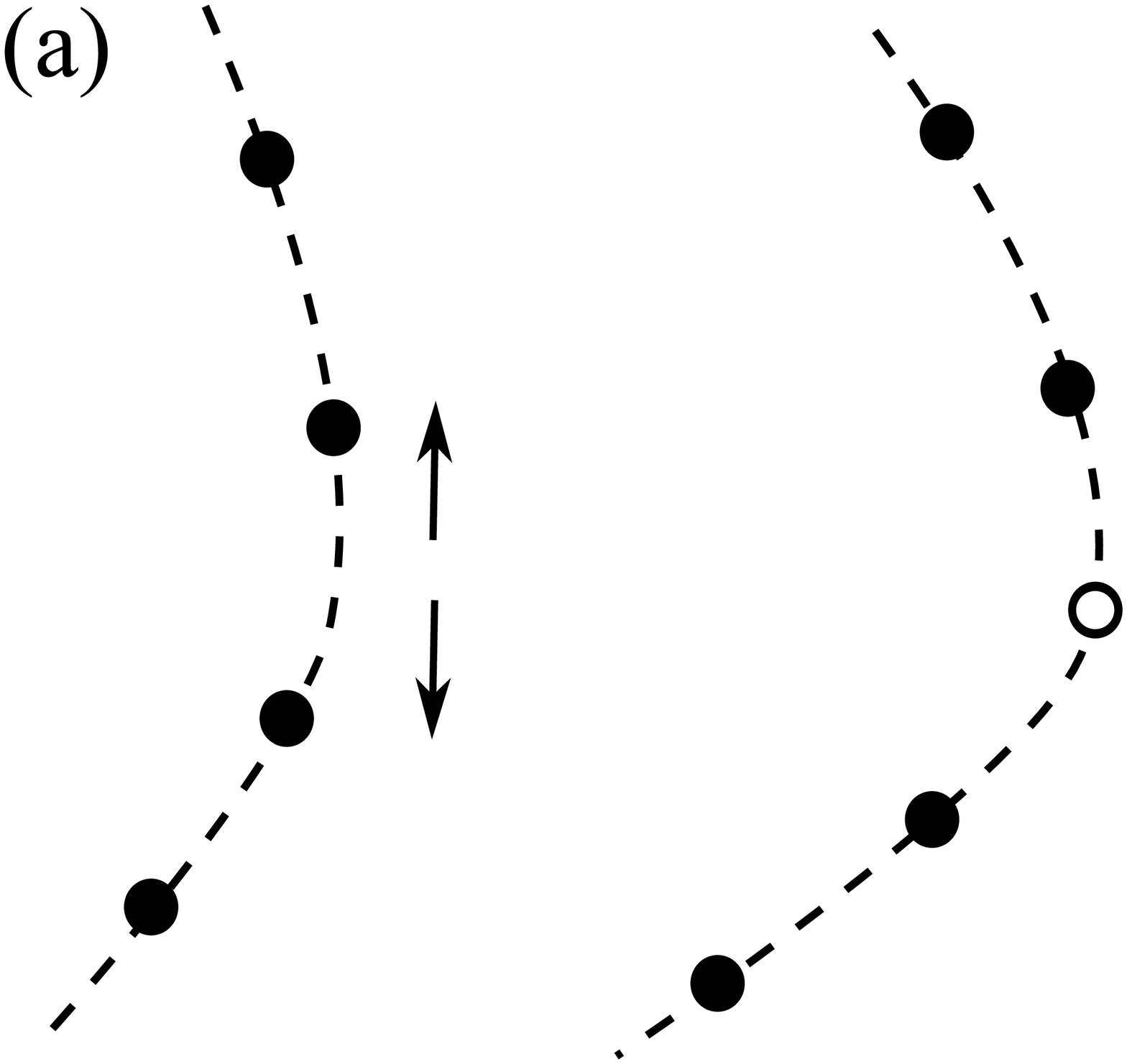}
    \hfill
    \includegraphics[width=0.4\textwidth]{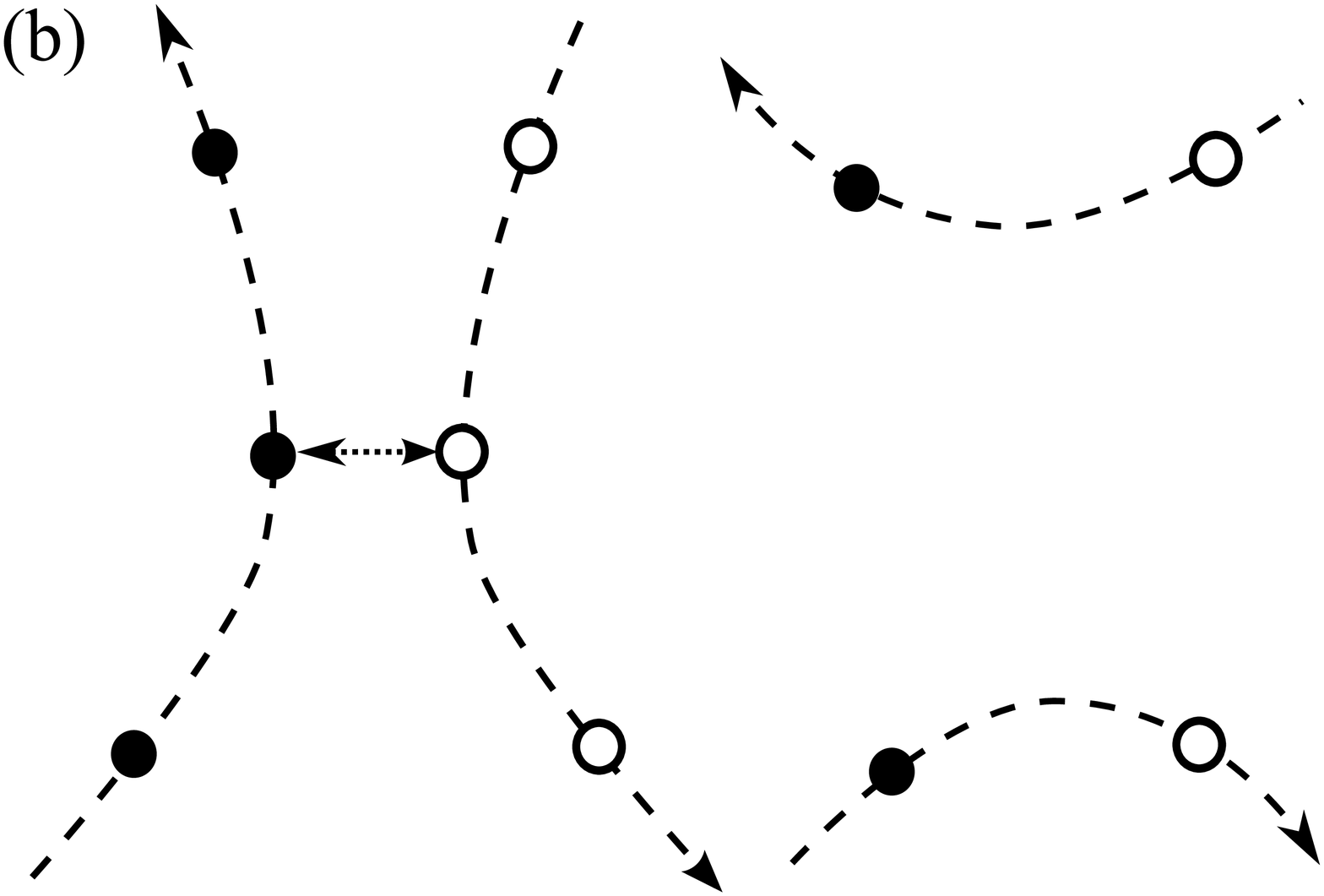}
    \caption{Schematic of the algorithms: (a) for the insertion of a new point (open circle) in a spreading front;  (b) for the removal of encroaching points (central circles) in merging fronts.}
\label{fig:schematics}
\end{center}
\end{figure*}

\begin{figure*}
\begin{center}
    \includegraphics[width=0.45\textwidth]{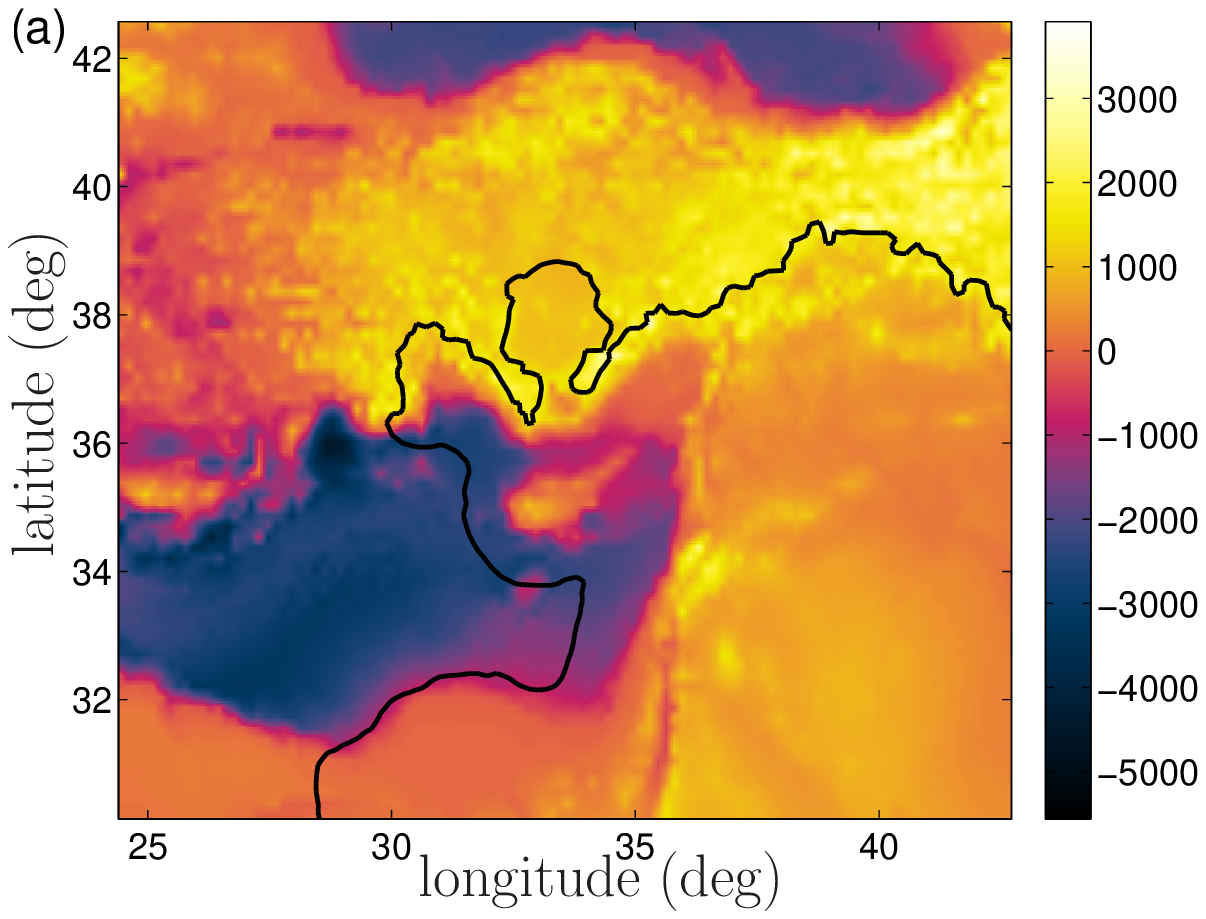}
\hfill
    \includegraphics[width=0.45\textwidth]{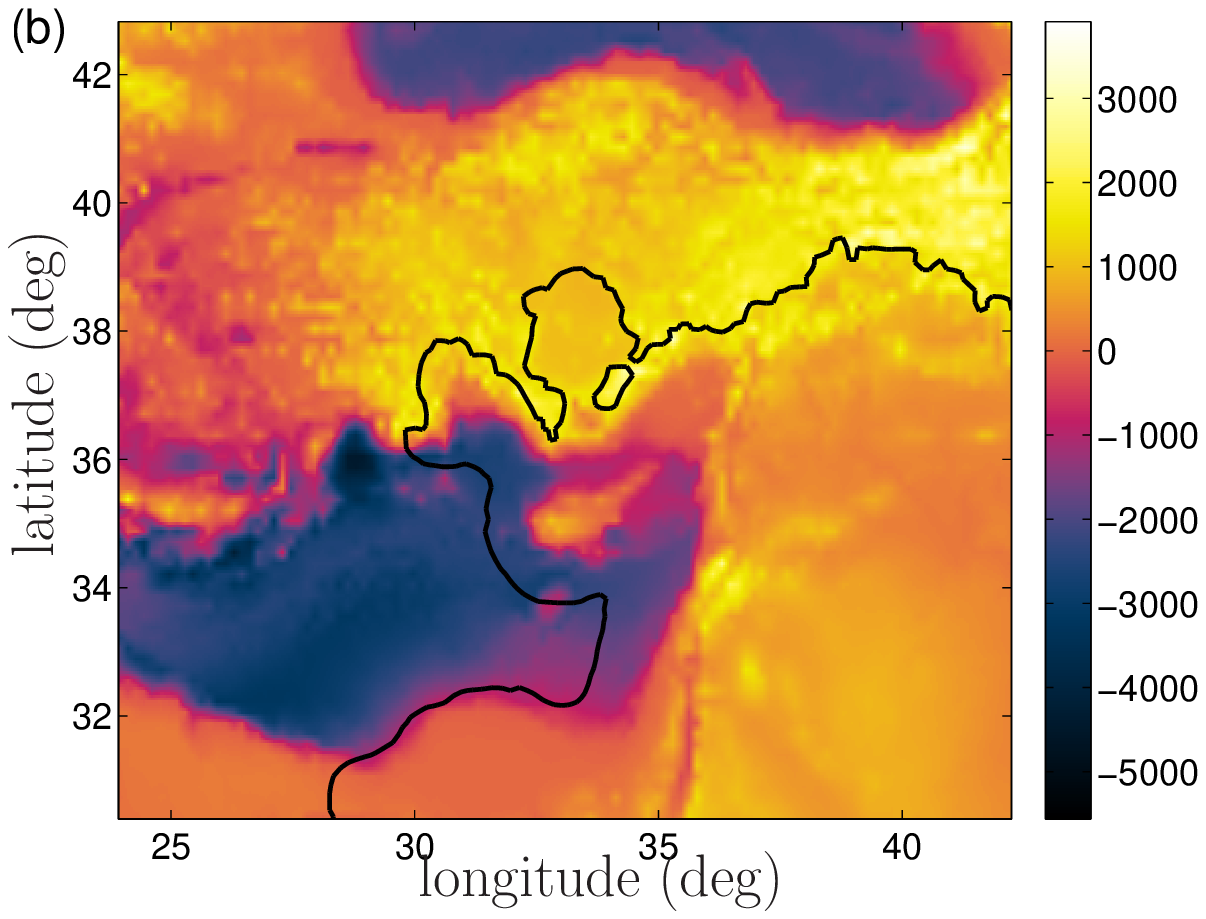}
    \caption{(Color online) The wavefront model plotted just before (a) and after (b) a merging event. The color gradient plot displays the altitude (in meters) for this region (the eastern Mediterranean). Note that the small loop shown behind the merged front would normally be removed from the calculation, but has been left in here for illustrative purposes.}
\label{fig:recon_snap}
\end{center}
\end{figure*}

We monitor the separation between the particles, and if this becomes larger than some specified value $\delta$, we introduce a new particle in order to maintain a roughly constant resolution along the front;
see Fig.~\ref{fig:schematics} (a) for a illustration of this process.
Due to the irregular nature of the velocity map shown in Fig.~\ref{fig:nuVR}, 
the wavefront can separate around low velocity regions 
(e.g.\ mountain ranges),
and subsequently re-merge.
We apply algorithms initially used to model the evolution of magnetic flux tubes in astrophysical simulations \cite{Baggaley:2009} 
to merge wavefronts in the particle model.
Every time-step we check the distance between each particle and all the other particles in the simulation.
If the separation of any two particles (who are not neighbors) is less than the resolution length $\delta$, then we
remove the encroaching points and
switch the ordering of the loops, so as to merge the fronts;
a schematic of this process is shown in Fig.~\ref{fig:schematics} (b).
Typically this process results in a merging of the main front and the creation of a small loop behind the main front, 
which we normally remove from the simulation to avoid unnecessary numerical effort.
Snapshots from a simulation where these small loops are not removed are shown in Fig.~\ref{fig:recon_snap}.
In the work reported here, we take $\delta$ to be equal to our grid spacing of
4 arc-minutes.


\subsection{Comparison between simulations}

We now present qualitative results on the comparison between the
wavefront model and the PDE model for a typical parameter set.  For
these (and subsequent) calculations, we place the initial source of
the farming population at $40^\circ$N, $35^\circ$E, within the Fertile
Crescent.  The starting time for the spread is taken as 6572 years cal
BC as this date is consistent with \citep{Davison:2006} and the
radiocarbon data at the site Tell Kashkashok.  The solution to the
FKPP equation, Eq.~(\ref{eq:FKPP}), is solved numerically,
approximating spatial derivatives using a second-order
finite-difference scheme, and time stepping using an Euler scheme.
The spatial resolution is 4 arc-seconds, on a 740 by 1100 mesh.  (This
is the same mesh introduced above to control the spatial variations
for our wavefront model; the same spatial variations are used for the
FKPP equation.)  The reduced computational effort of the wavefront
method provides us with a significant speed-up, with a typical
simulation taking approximately 10 seconds on a single processor with
a clock speed 2.67GHz.  The corresponding solution to the FKPP
equation, with a time-step satisfying the Courant-Friedrichs-Lewy
(CFL) condition \cite{Courant:1967}, requires approximately 24 hours
on an eight-processor cluster with hyper-threaded Intel Xeon quadcore
processors of the same clock speed.

\begin{figure*}
\begin{center}
    \includegraphics[width=0.48\textwidth]{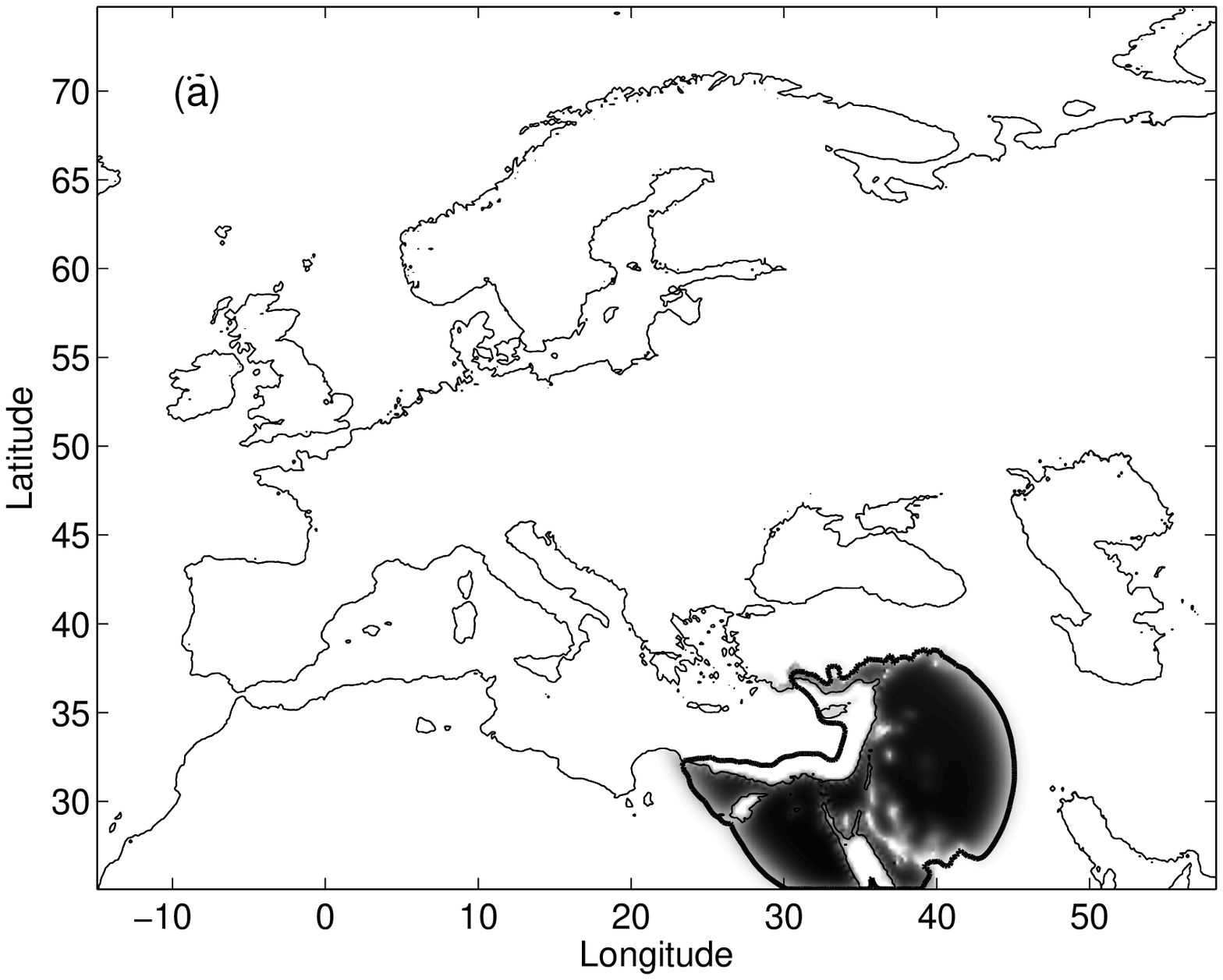}
\hfill
    \includegraphics[width=0.48\textwidth]{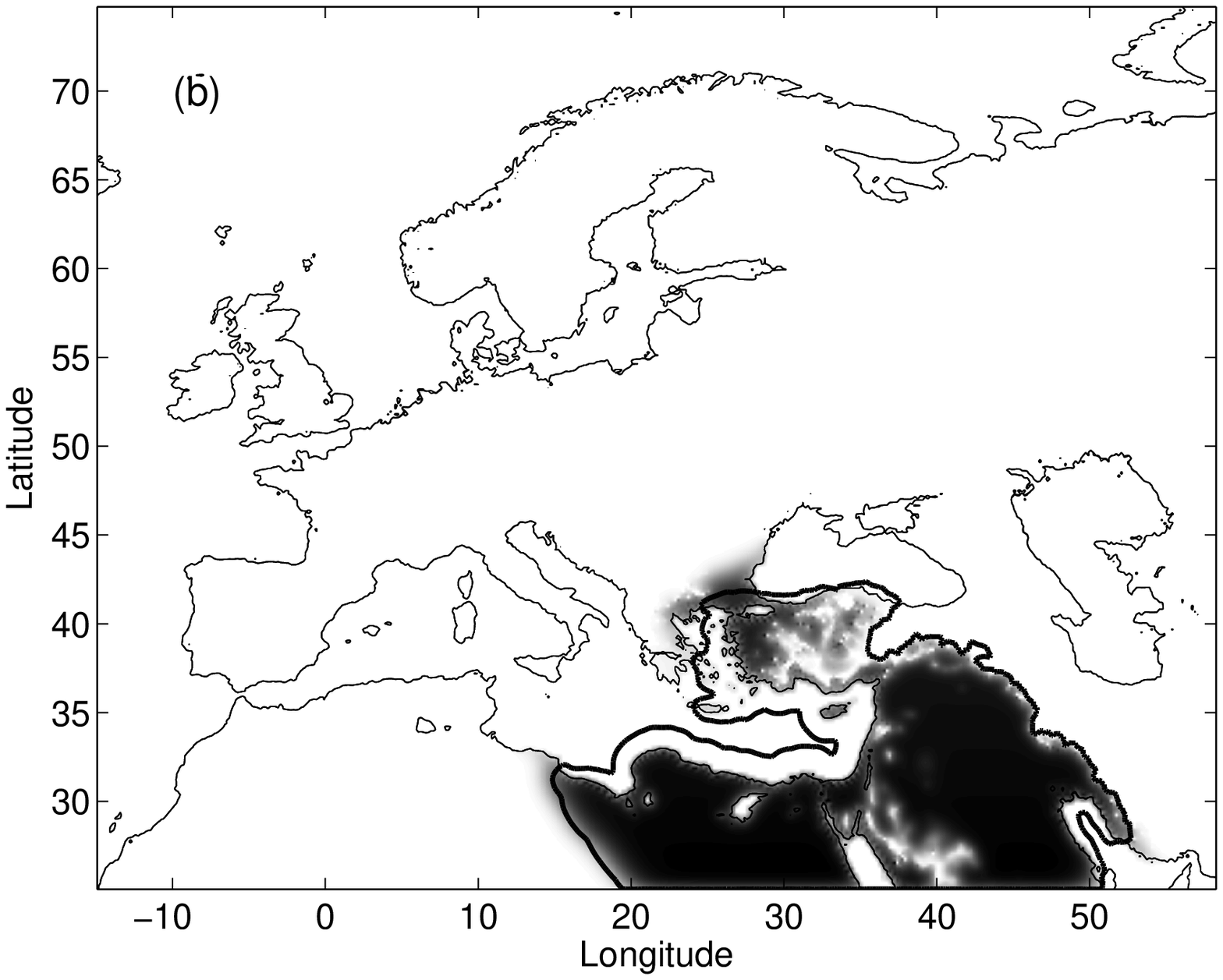}\\
    \includegraphics[width=0.48\textwidth]{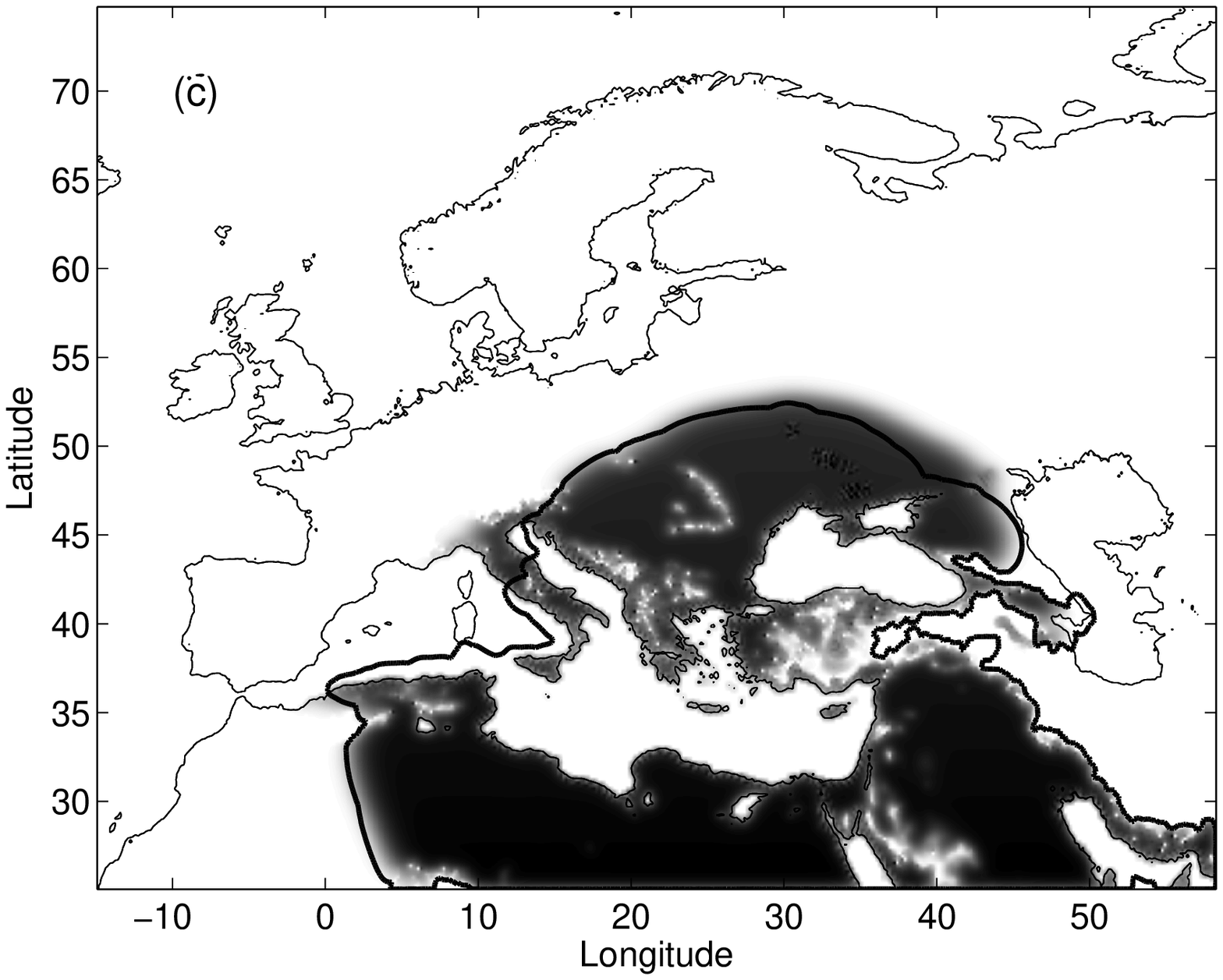}
\hfill
    \includegraphics[width=0.48\textwidth]{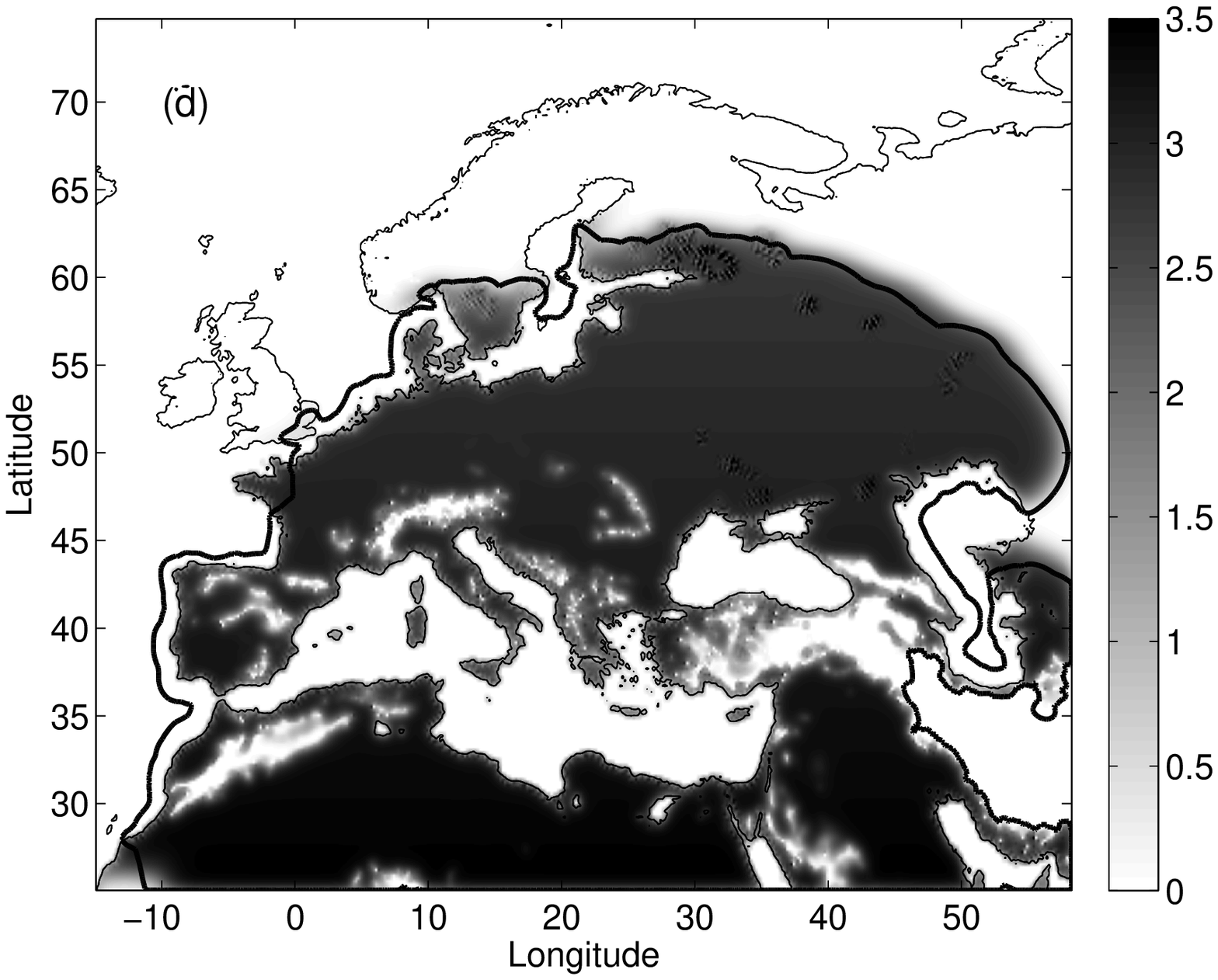}
    \caption{A comparison between the numerical solution to the FKPP equation and our propagating wavefront model. The output from the wavefront model is shown by the thick black lines.
The corresponding output from the FKPP model is shown by the grayscale plots of the population density, with the scale given to the right of panel (d); the values are in terms of people per square kilometer.
The four panels are for times 1000 years (a), 1800 years (b), 3000 years (c), and 4600 years (d) after the start of the simulations. In the wavefront simulations $U_0=1$km/year, $V_{\rm C}=2$km/year, $V_{\rm R}=0$. A consistent parameter set is used in the FKPP simulation.}
\label{fig:comparison}
\end{center}
\end{figure*}

Whilst we do not expect an exact agreement between the two models, 
we do expect their arrival time at a particular site to be similar.
Indeed, we do find a reasonable agreement between the two models,
with a maximum discrepancy in the arrival time of approximately 120 years and typical values less than 50 years (for a simulation covering of order 5000 years).
Fig.~\ref{fig:comparison} shows snapshots at four times during the simulations.


\section{Radiocarbon Data}\label{sec:rcarb}

\begin{figure}
\begin{center}
    \includegraphics[width=0.48\textwidth]{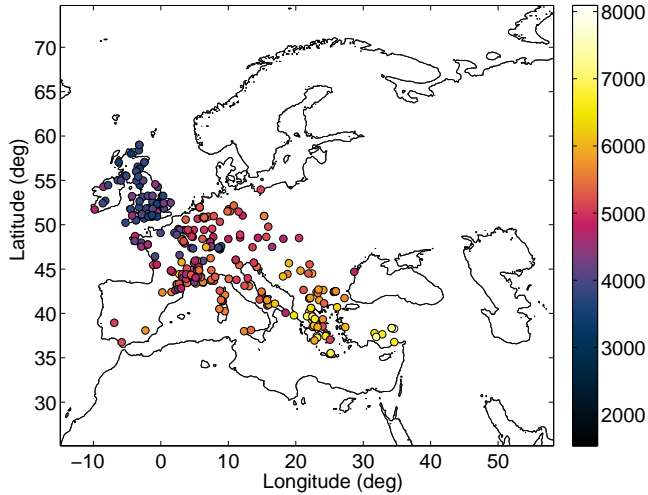}
    \caption{The radiocarbon sites we use to constrain our simulations. The colorbar shows the estimated first arrival time at a site, in time BC, from the radiometric data.}
\label{fig:rcarb}
\end{center}
\end{figure}

We use a compilation of 302 dates from sites in Southern and Western
Europe from \cite{Gkiasta:2003}, \cite{Shennan:2000} and
\cite{Thissen:2006}. These data contain multiple dates per site and so
we determine a single date for each site by using a method based on
\cite{Dolukhanov:2005}. The method we use is described in detail in
Davison \etal \cite{Davison:2007}. Briefly, for sites with at least
eight date measurements, a $\chi^2$ statistical test is used to
determine the most likely first arrival date from a coeval sub-sample,
and for sites with fewer measurements, we use a weighted mean of these
measurements.  Fig.~\ref{fig:rcarb} shows a plot of the radiocarbon
sites shaded according to the estimated first arrival time, $t_i$,
obtained from this statistical treatment.

In this paper we focus on a statistical model with a single error term
accounting for mismatch between the data and the wavefront; this is
described in the next section.  Whilst this approach is not entirely
satisfactory, the additional complication of properly accounting for
varying site errors would add another level of complexity, which was
deemed prohibitive for this initial investigation. However, a
statistical model which accounts for errors in the dates that can vary
between sites may be adopted in future studies.


\section{Bayesian Inference}\label{sec:MCMC}
We now outline the proposed statistical model and provide details of
our Bayesian inference scheme.

Let $\tau(\mathbf{x}_i|\boldsymbol{\theta})$ denote the time at which
the wavefront arrives at site $i$, at position $\mathbf{x}_{i}$, for
$i=1,\ldots,n$, where $n=302$ is the number of radiocarbon sites in
our data set.  Here $\boldsymbol{\theta}=(U_0, V_{\rm C}, V_{\rm
  R})^T$ is the vector of model parameters about which we want to draw
inferences.  The observed arrival time at site~$i$ is denoted by $t_i$
and these times are displayed in Fig.~\ref{fig:rcarb}. Our statistical
model assumes that these data are generated by the wavefront model
subject to (spatially) independent normal errors. Specifically, at
site~$i$ we have
\begin{equation}\label{eq:stats_model}
t_i=\tau(\mathbf{x}_i|\boldsymbol{\theta})+\sigma \epsilon_i,
\quad i=1,\ldots,n,
\end{equation}
where the $\epsilon_i$ are independent and identically distributed
standard normal random variables and $\sigma$ is the spatially
homogeneous standard deviation, allowing for a mismatch between the
model and the observations (i.e.~local variations in the arrival of
the wavefront, corresponding to the expected local deviations from the
idealized, global model).

By adopting a Bayesian approach to the problem of inferring the model parameters, we 
express initial beliefs about likely parameter values via 
a prior distribution, 
denoted $\pi(\boldsymbol{\theta},\sigma)$.
We then construct the posterior distribution 
of our parameters, given the observed arrival times.
Bayes' theorem gives this posterior distribution as
\begin{equation}\label{eq:bayes}
  \pi(\boldsymbol{\theta},\sigma|\boldsymbol{t}) \propto \pi(\boldsymbol{\theta},\sigma)
\pi(\boldsymbol{t}|\boldsymbol{\theta},\sigma),
\end{equation}
where $\pi(\boldsymbol{t}|\boldsymbol{\theta},\sigma)$ is the likelihood function,
i.e.\ the joint probability of the observed arrival times,
regarded as a function of the parameter values. 
If we assume the model outlined in Eq.~(\ref{eq:stats_model}),
then we can write the likelihood function as
\begin{equation}\label{eq:likelihood1}
\pi(\boldsymbol{t}|\boldsymbol{\theta},\sigma) 
\propto \sigma^{-n}\displaystyle 
\exp{\left[ -\frac{1}{2\sigma^2} \sum_{i=1}^{n} \{t_i-\tau(\mathbf{x}_i|\boldsymbol{\theta})\}^2 \right]}.
\end{equation}
We specify our fairly weak \emph{a priori} beliefs about
$\boldsymbol{\theta}$ by adopting independent lognormal distributions
for the components $U_0$, $V_{\rm C}$ and $V_{\rm R}$, with modes
chosen to match previous archaeological estimates
\cite{Ammerman:1971,Gkiasta:2003,Zilhao:2003}.  Specifically we take
$U_0\sim\textrm{LN}(0.5,0.71^2)$, $V_{\rm C}\sim\textrm{LN}(1,0.5^2)$
and $V_{\rm R}\sim\textrm{LN}(2.2,0.8^2)$.  We use a weakly
informative inverse Gamma prior for the global error parameter, with
$\sigma^2 \sim IG(5,10^6)$.

Due to the complex dependence of the wavefront solution on the model
parameters, the posterior in Eq.~(\ref{eq:bayes}) is analytically
intractable.  Markov chain Monte Carlo (MCMC) methods are commonly
used, in the context of Bayesian inference, to sample intractable
posterior distributions.  These methods aim to construct a Markov
chain whose invariant distribution is the desired posterior
distribution.  Such approaches are particularly useful for Bayesian
inference since the target distribution need only be known up to
proportionality.  A recent review of these methods can be found in
\cite{gamerman2006markov}.

In this paper we focus on a Gibbs sampler \cite{geman1984}. This particular 
MCMC scheme can be useful for sampling from high dimensional distributions,
and requires the ability to sample from the full conditional distribution 
of each parameter (or more generally, subsets of parameter components). 
In the absence of analytically tractable full conditionals, a 
Metropolis-Hastings scheme can be used for this. Such an 
approach is often termed \emph{Metropolis within Gibbs}, and 
its use is outlined in the next section. 

\subsection{Markov chain Monte Carlo algorithm}\label{subsec:MCMC}

In this section we provide details pertinent to our implementation of
the MCMC scheme.  A more detailed description of the algorithm can be
found in Appendix~\ref{app:met_hast}. 

We consider a Gibbs sampling strategy where we alternate between draws
of $\boldsymbol{\theta}$ and draws of $\sigma^2$ (and therefore
$\sigma$) from their full conditional distributions. The
form of the statistical model and its inverse Gamma prior permit an
analytically tractable full conditional for~$\sigma^2$. Consequently
realizations of $\sigma$ can be sampled directly. The full conditional
density of $\boldsymbol{\theta}$, namely
$\pi(\boldsymbol{\theta}|\sigma,\boldsymbol{t})$, however, is
intractable and we therefore use a Metropolis-Hastings scheme to
sample from the corresponding distribution.  In brief, a Markov chain
is constructed by generating candidate values of each component of
$\boldsymbol{\theta}$ via a symmetric random walk with normal
innovations on the log-scale: this ensures proposed parameter draws
are non-negative. A proposed value is accepted as the next value in
the chain with a probability that ensures the Markov chain has
invariant distribution given by the distribution of interest.  If a
proposal is not accepted then the next value for that parameter is
taken to be its current value.  The acceptance probability requires
that the target density can be evaluated up to proportionality. Each
MCMC iteration therefore requires a single run of the wavefront model
expanding across the whole of Europe.

Unfortunately, the number of MCMC iterations required to produce near
independent draws from the joint posterior distribution precludes
using the wavefront model to evaluate each
$\tau(\mathbf{x}_i|\boldsymbol{\theta})$.  (The individual wavefront
model calculations run too slowly, given the large number of
iterations required.) 

To proceed, we seek a faster approximation of the first arrival times
from the wavefront model. One option is to use a deterministic approximation, such as linear interpolation or cubic splines. Initially the wavefront model would be run
at a specific set of parameter values, and the arrival time at each site stored.
Parameters within the interpolation method could readily be computed from this output.
The arrival time at new parameter sets could then be approximated using the chosen deterministic `emulator'.
In the statistical literature \cite{kennedy01}  Gaussian process emulation is favored, as
this not only interpolates smoothly between design points but also quantifies levels of
uncertainty around interpolated values.
Further details on how to build and test such emulators can be found in
Appendix~\ref{app:emulator}. Using these emulators to approximate the
first arrival times $\tau(\mathbf{x}_{i}|\boldsymbol{\theta})$ at each
site makes the MCMC scheme outlined above computationally practicable.
The scheme produces a sample from the joint posterior distribution of
our model parameters.

\section{Results}\label{sec:results}

We now present results obtained from the output of the MCMC scheme.
We performed $5.5 \times 10^6$ iterations of the algorithm before
discarding the first $5 \times 10^5$ parameter draws as `burn in' to
allow the chain to converge.  The remaining $5\times 10^6$ iterates
were then thinned to reduce the autocorrelation in the sample: we took
every $500^\textrm{th}$ iterate, leaving a sample of $10^4$ (almost)
uncorrelated values from the joint posterior distribution.  We
assessed convergence of the MCMC scheme by repeating the above
procedure for many different starting parameter sets (randomly drawn
from the prior distribution) and found no problem with convergence.

\begin{figure*}
\begin{center}
    \includegraphics[width=0.245\textwidth]{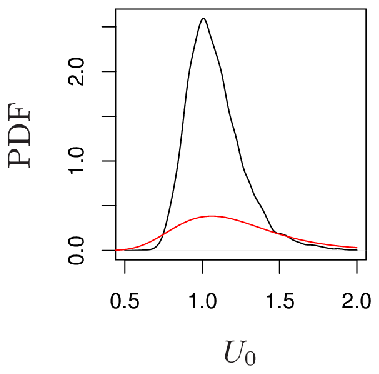}
    \includegraphics[width=0.245\textwidth]{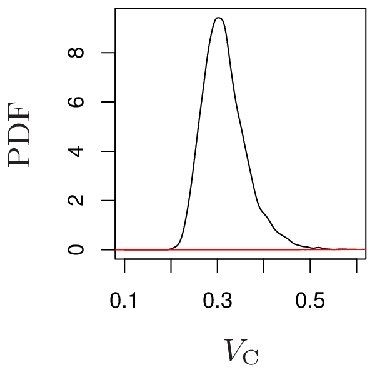}
    \includegraphics[width=0.245\textwidth]{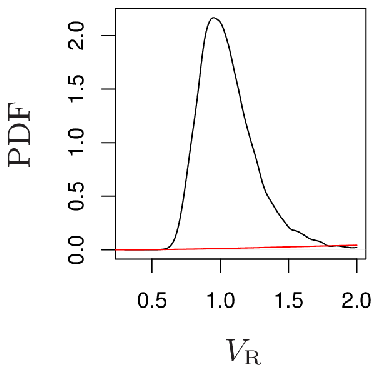}
    \includegraphics[width=0.245\textwidth]{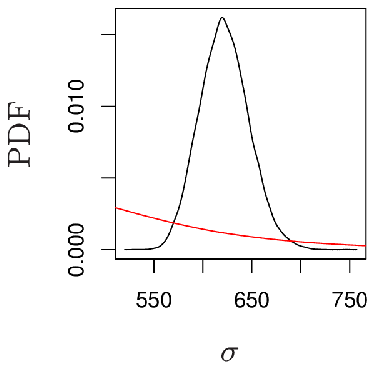}
    \caption{(Color online) Marginal posterior densities (in black; prior
      distribution in red) for (a) $U_0$, (b) $V_{\rm C}$, (c) $V_{\rm R}$ and
      (d) $\sigma$, based on the (thinned) output of the MCMC scheme,
      using a Metropolis within Gibbs sampler.}
\label{fig:param_chains}
\end{center}
\end{figure*}

The output of the MCMC scheme is summarized in
Fig.~\ref{fig:param_chains}. It shows kernel density estimates
\cite{Silverman:1986} of the marginal posterior probability density
function, for each of $U_0$, $V_{\rm C}$, $V_{\rm R}$ and $\sigma$. 
For the three parameters in our mathematical model, the modes of the
marginal posterior distributions (in black in
Fig.~\ref{fig:param_chains}) are of the magnitude expected from other
studies in the literature (as described in section~\ref{sect:intro}).
Compared with their respective prior distributions (in red), the
posterior distributions are considerably tighter, showing that the
radiocarbon data have indeed been informative, and have effectively
constrained the plausible range of model parameters.  For example,
posterior samples of the background wavespeed, $U_0$ (with a mode of
approximately 1 km/year, and a 95\% range of 0.79--1.41 km/year), are
in good agreement with the studies cited in section~\ref{sect:intro}
\citep{Ammerman:1971,Ackland:2007,Davison:2006}.  In terms of an FKPP
model, with a growth rate of $\gamma\simeq0.02\,\textrm{year}^{-1}$
(of the order typically used in such models), this would correspond to a
diffusivity of $\nu\simeq13\,\textrm{km}^2/\textrm{year}$; this value
is also comparable to those typically used in FKPP models.

The only previous studies which have modeled 
an enhanced population mobility along waterways 
\citep{Davison:2006,Davison:2009}
were motivated by studies based on specific local phenomena. 
For example, the incorporation of enhanced coastal speeds 
within the model was motivated by radiocarbon evidence for the
spread of the Impressed Ware culture along the coastline of the
Western Mediterranean,
with some estimates of speeds of order 10\,km/year 
\citep{Zilhao:2001,Zilhao:2003}.  
Whilst some form of enhanced spread along the coastline of this region
may be required,
extrapolating to a similar spread along all of Europe's coasts 
might very well give an inferior fit to the data as a whole. 
(E.g., if we take the rate of spread along all of
Europe's coastlines to be of order 10\,km/year,
then the wavefront may arrive at many sites 
far earlier than the radiocarbon data suggest.)
The inference presented in this paper seems to confirm this,
as marginal posterior samples of the global coastal propagation speed,
$V_{\rm C}$
(with a mode of around $0.3$km/year,
and a 95\% range of 0.23--0.41 km/year)
are markedly lower 
than the values quoted for the Impressed Ware culture.

Posterior samples of $V_{\rm C}$ are also significantly lower 
than the corresponding value used in \cite{Davison:2006} (20\,km/year).
It may be that, compared with that earlier work, 
our more rigorous method of comparing models against the data 
suggests an improved model fit without a large enhanced coastal speed,
and with regional data variations 
(such as those associated with the Impressed Ware culture)
largely accounted for within the global error parameter in our statistical
model (discussed further below).
This may not entirely explain the difference between the two studies, however;
it is also possible that the implementation of coastal advection 
in \cite{Davison:2006} has exaggerated the overall magnitude
of this effect.
The value quoted there for the advection actually only applies at points
exactly on the coast, whereas nearby locations experience a reduced velocity,
decreasing with their distance from the coast;
as a result, the mean, effective, advective speed
may be somewhat lower than the peak value quoted.

In terms of the river advection, the marginal posterior values of
$V_{\rm R}$ are clearly non-negligible (with a mode of approximately
1.0 km/year, and a 95\% range of 0.72--1.38 km/year); it is comparable
to the background spread modeled by $U_{0}$, vindicating the
suggestion of anisotropic spread along these river basins.  However,
this value is significantly smaller than the values normally quoted
for the spread of the LBK culture (of order 5\,km/year)
\cite{Davison:2006}.  This deviation may be due to the reasons discussed 
above for the
coastal velocity. However, the discrepancy with the widely-accepted
archaeological timescale for the LBK culture (which does not refer to
a spatial mathematical model, but is obtained directly from a coeval
set of radiocarbon dates) requires further explanation.  Looking in
detail at the data in this region, it may be that the earliest dates
associated with the LBK culture simply do not correspond well to
spread by a continuous wave of advance (anisotropic or otherwise);
rather, early settlements may have been been formed by something like
a leapfrog or pioneer mechanism, and the whole region only settled
(and outward spread continued) after some subsequent delay. In this
paper we have studied a model of large-scale spread, but it may be
the model is too crude and would need small-scale refinement to
explain the spread of the LBK culture.

For the global error parameter in our statistical model
(Eq.~\ref{eq:stats_model}), $\sigma$, the marginal posterior
distribution is centred on a value of order 600 years (see
Fig.~\ref{fig:param_chains}).  The 95\% confidence range is 577--671
years.  This is significantly larger than the values typically quoted
for the effective minimum uncertainty of radiocarbon dates for this
period, of order 160 years \citep{Dolukhanov:2005}.  (The latter value
is derived empirically from well explored, archaeologically
homogeneous sites, effectively allowing for sample contamination and
other sources of errors; this may be contrasted with the quoted
laboratory uncertainties, which characterize only the accuracy of the
laboratory measurement, regardless of the provenance of the sample.)
The global parameter ($\sigma$) in our model, however, does not merely
reflect the uncertainty in the dating, but also allows for the
mismatch between our mathematical model of the spread (a globally
continuous wave of advance) and the regional variations present in the
actual spread.  Thus our inference suggest that a simple global wave
of advance across Southern and Western Europe, while remaining a good
model on the continental scale, should only be considered a good model
on timescales of order 600 years (and thus lengthscales of order 600
km) or greater; on shorter timescales (and lengthscales), significant
local variations should be expected.  As noted briefly above (in our
discussion of the advective velocities), this parameter within our
model may to some extent allow for regional variations that might
alternatively be modeled by specific regional effects (e.g.\ river
advection), thus explaining the relatively low values of our inferred
advective speeds.  The posterior distributions suggest that this type
of fit --- requiring relatively large global uncertainty, but then
favoring relatively low local advective speeds --- is the optimum way
of explaining our dataset of first arrival times within a wave of
advance model of the sort presented here.  Of course, the conclusions
of this inference may depend upon specific features of the models used
here (both mathematical and statistical), and may vary for different
models.  In extensions to the current work we will explore alternative
models, e.g.\ allowing for increased regional variation in the coastal
advection, and allowing for spatial correlation between nearby sites
within the statistical model.  We will also consider additional, more
recent, radiocarbon data, which will provide observed arrival times at
new sites, and will thereby allow an out-of-sample assessment of
prediction error.

\begin{figure*}
\begin{center}
    \includegraphics[width=0.325\textwidth]{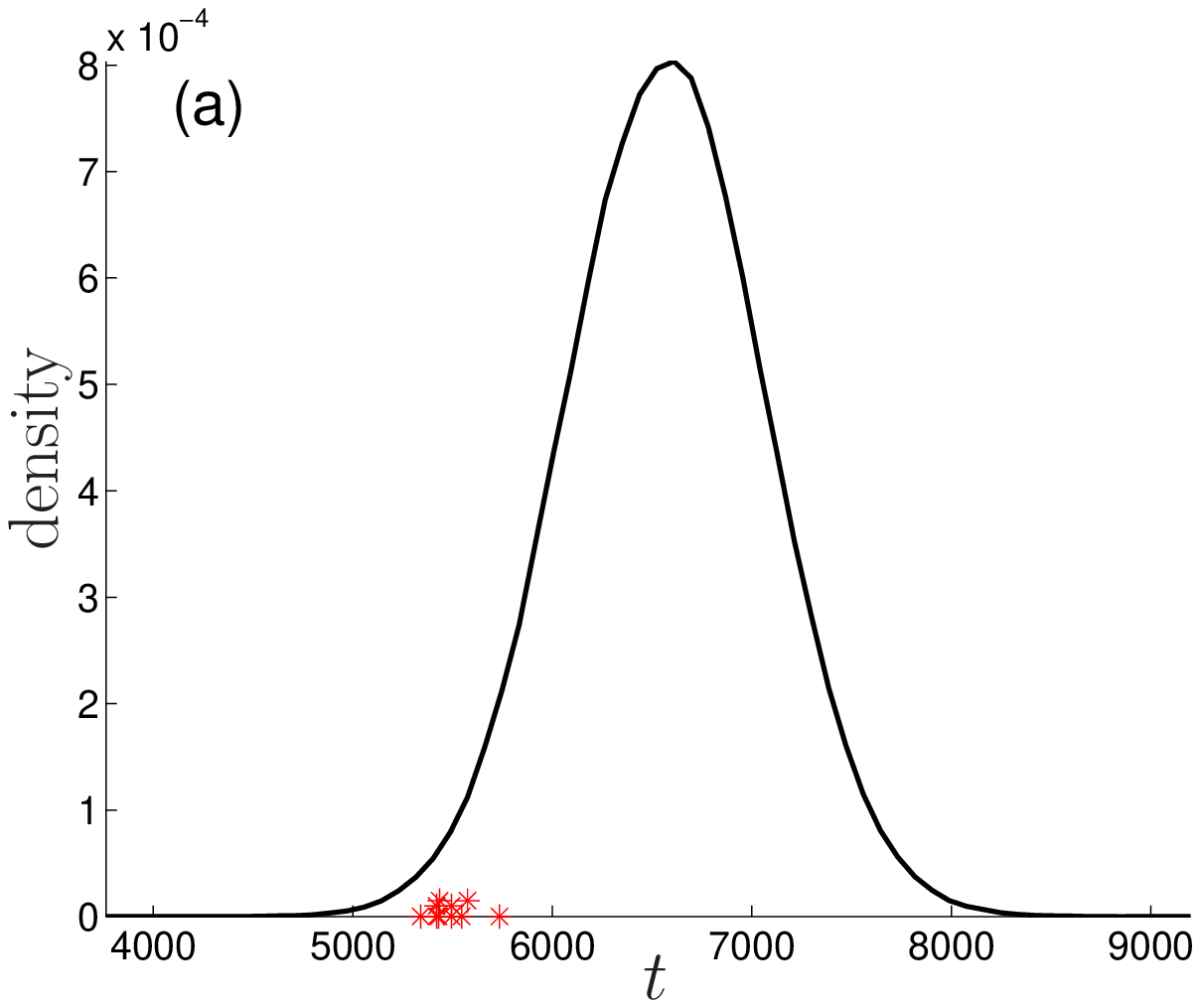}
\hfill
    \includegraphics[width=0.325\textwidth]{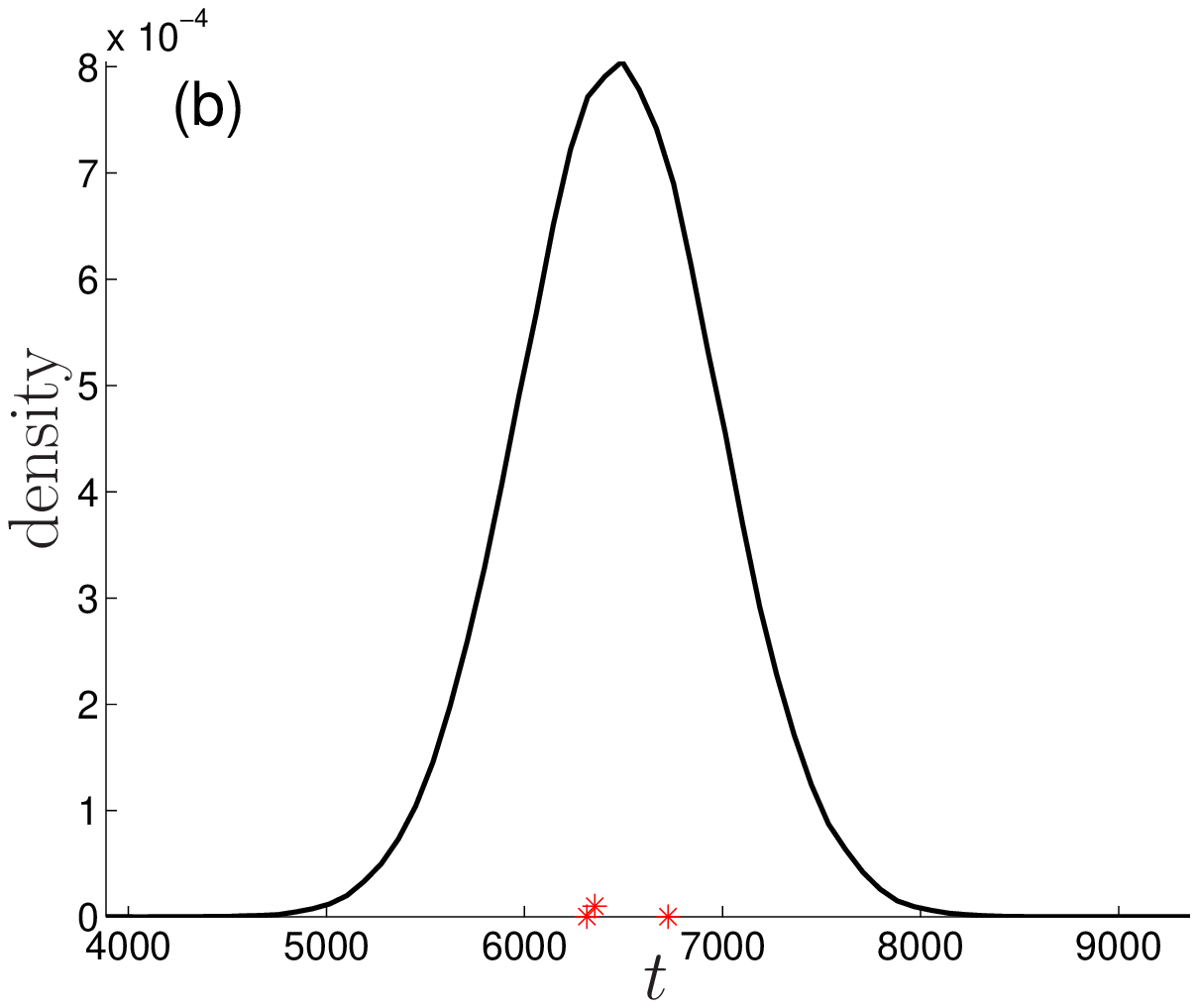}
\hfill
    \includegraphics[width=0.325\textwidth]{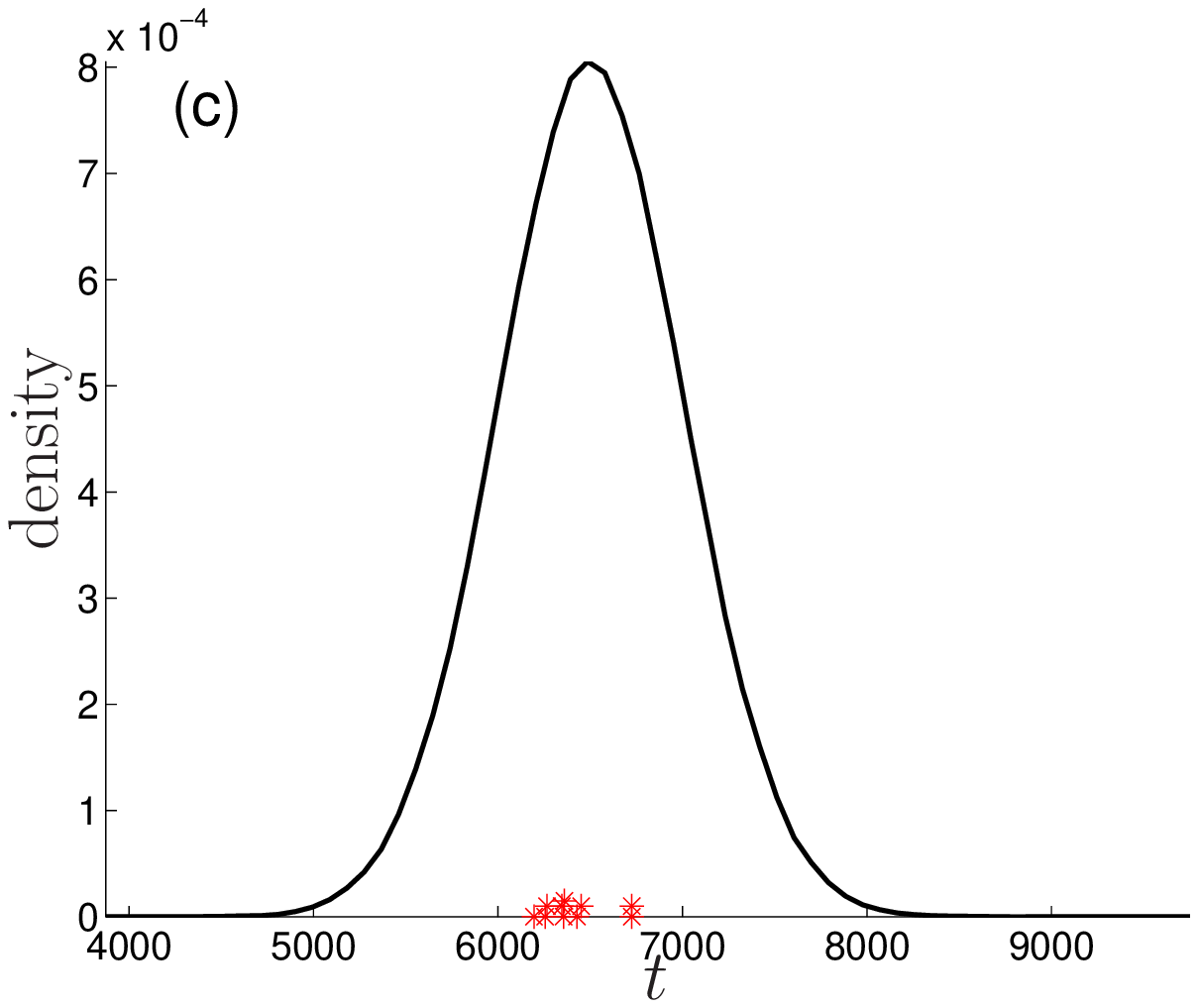}
    \caption{(Color online) Predictive densities for the arrival time at three sites
      plotted as black lines, with the observed radiocarbon dates plotted as (red)
      asterisks. (a) Site Kremenik, located at
42.3N, 23.27E, an example of a bad model fit, where our model
predicts a much earlier arrival time than is presently observed; this
point can be seen plotted as a triangle in
Fig.~\ref{fig:globalcoloured}. The middle and right panels are examples
of more typical agreement between the model and data; (b)
Agrissa Magoula (39.63N, 22.47E), (c) Seskto (39.28N, 22.82E).}
\label{fig:predictive_comp}
\end{center}
\end{figure*}

In addition to performing inference for the model parameters, we 
use predictive simulations to assess the validity of the statistical 
model and the underlying model of the wave of advance. The 
posterior predictive distribution of the arrival time at a site $i$, 
$t_{i,\textrm{pred}}$, can be determined as follows. 
Using Eq.~(\ref{eq:stats_model}) we have that
\[
t_{i,\textrm{pred}}|\boldsymbol{\theta},\sigma\sim
\textrm{N}\left(\tau(\mathbf{x}_i|\boldsymbol{\theta}),\sigma^{2}\right),
\]
where $\tau(\mathbf{x}_i|\boldsymbol{\theta})$ is the arrival time of
the wavefront, approximated by the emulator. We therefore take each
sampled parameter value $(\boldsymbol{\theta}^{(j)},\sigma^{(j)})$
from the MCMC output and generate a realization from the predictive
arrival time distribution by simulating realizations from
$t^{(j)}_{i,\textrm{pred}}|\boldsymbol{\theta}^{(j)},\sigma^{(j)}$.
We thus obtain a sample of first arrival times at each site.
Fig.~\ref{fig:predictive_comp} shows the predictive densities for
three sites.

\begin{figure*}
\begin{center}
    \includegraphics[width=0.6\textwidth]{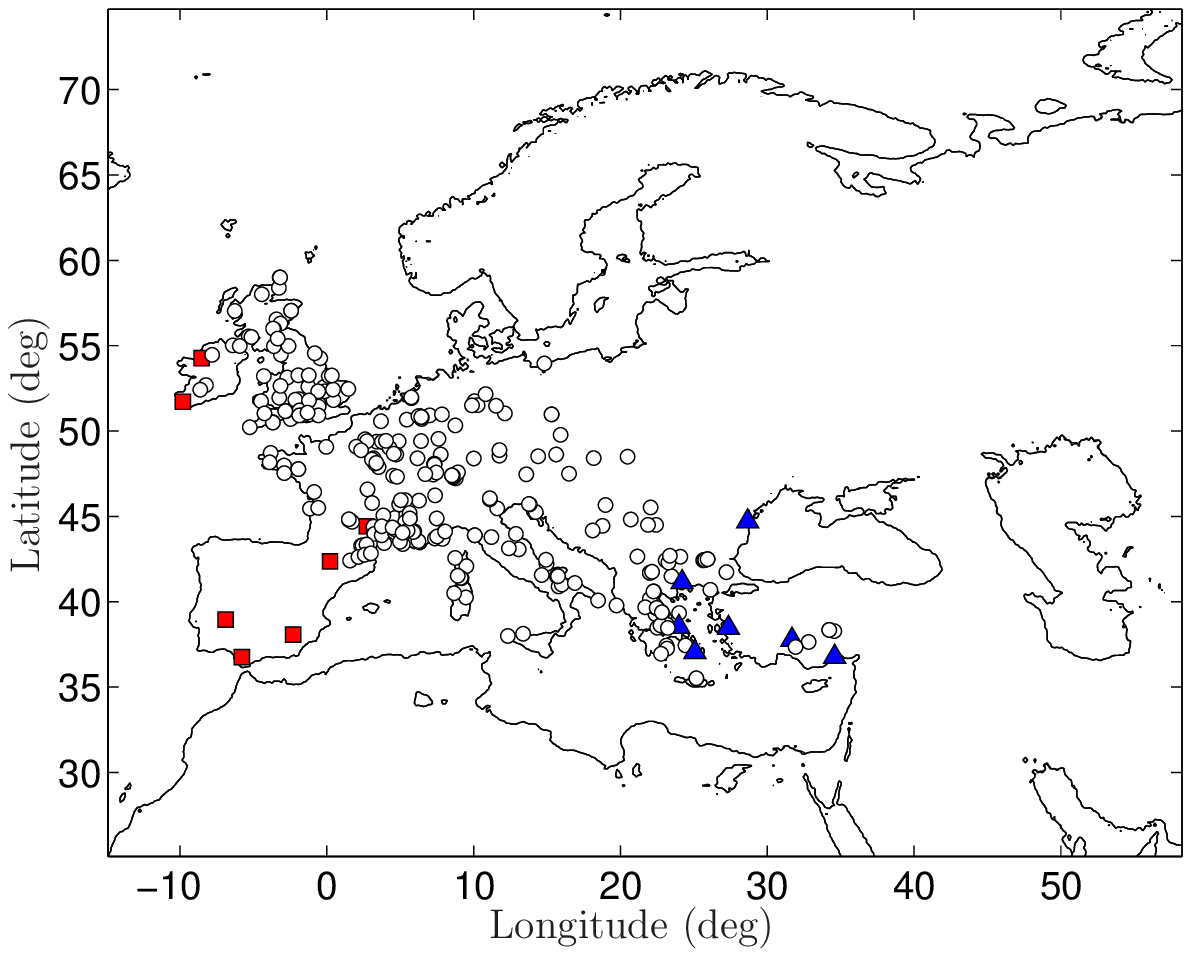}
    \caption{(Color online) Discrepancies between the predicted first arrival time
      (modal values of $t_{i,\textrm{pred}}$) and the observed values
      ($t_{i}$).  Filled blue triangles (red squares) show sites where
      the model predicts anomalously early (late) arrivals. Open
      circles show sites with more acceptable agreement.}
\label{fig:globalcoloured}
\end{center}
\end{figure*}

To see where our model is failing to agree with the radiocarbon data,
sites where the observed radiocarbon date 
falls outside an approximate
95\% credible interval for predicted first arrival 
are plotted as filled symbols in Fig.~\ref{fig:globalcoloured};
sites which fall inside this interval are plotted as open circles.
We distinguish between sites where our model predicts an earlier arrival time than is observed in the radiocarbon data, 
and those where our model arrives late.
Where we predict an earlier arrival time, 
it is quite possible that the radiocarbon data at the site 
are simply from a relatively late settlement within this local region,
and earlier data there have yet to be discovered.
Where the model predicts a later first arrival time than is observed,
then this may be an indication that some localized process,
which we have not included in our model,
has caused a much faster spread in this region.


\section{Conclusions}
\label{sec:conc}

In this paper we have introduced an innovative wavefront model, which
allows the efficient simulation of the spread of a wave of advance
model (with both isotropic and localized anisotropic components of
spread); we have applied this model to the spread of Neolithic culture
across Europe (with the localized anisotropy being associated with a
hypothesized enhanced rate of spread along certain waterways).  We
adopted a Bayesian approach to the problem of inferring the model
parameters given observed arrival times, which we assumed were given
by the wavefront model but subject to Gaussian error. A Markov chain
Monte Carlo scheme was used to sample the intractable posterior
distribution of the model parameters. To alleviate computational cost,
we constructed Gaussian process emulators for the arrival time of the
wavefront at each radiocarbon site. As a result, we obtain the
marginal posterior probability distributions for the model parameters
of interest: the background rate of spread ($U_{0}$), and the enhanced
rates of spread associated with coastlines ($V_{\rm C}$) and with the
Danube--Rhine river systems ($V_{\rm R}$).  To our knowledge, this is
the first attempt to apply such inference techniques to this problem.

We find that the posterior variance is reduced (relative to the prior 
variance) suggesting that the data have been informative.
Marginal posterior samples of $U_{0}$,
with a modal value of order 1 km/year,
are consistent with previous studies \citep{Ackland:2007,Davison:2006}.
Modal values for $V_{\rm C}$ and $V_{\rm R}$
are of order 0.3 km/year and 1 km/year,
respectively.  This value for the river advection ($V_{\rm R}\simeq
1\,$km/year) is clearly comparable to the speed of the background
spread ($U_{0}$), confirming that an enhanced spread within these
river basins can be robustly concluded from the data.  This value is
nevertheless significantly smaller than the value of 5 km/year often
quoted for the rate of spread of the local Neolithic culture (the LBK
culture) \cite{Zilhao:2001,Dolukhanov:2005}.  A closer inspection of
the relevant data suggests that the spread of this particular culture
may not be particularly well modeled by a continuous wave of advance,
and subsequent models for this region may wish to pursue other
possibilities; the estimate of 1 km/year given above should simply be
considered as the best-fitting value within the constraints of the
current model.
The relatively low modal value for the coastal advection ($V_{\rm
  C}\simeq 0.3\,$km/year) suggests that such an advection, while not
negligible, should not be considered particularly significant
throughout Europe as a whole.  This is perhaps not surprising, given
that the principal motivation for this effect only applies to a
specific region of Europe (the Western Mediterranean coastline, along
which the Impressed Ware culture spread
\cite{Zilhao:2001,Zilhao:2003}).

In addition to performing inference for the parameters characterising
the wavefront, we also infer the `global error' $\sigma$ (here
formally introduced within our statistical model), representing both
uncertainty in the radiocarbon dates and the misfit between our simple
global wavefront model and the true spread (with its regional
variations and local anomalies).  The posterior modal and mean values
for $\sigma$ are of order of 600 years,
significantly larger than the uncertainty normally associated with
radiocarbon dates for sites of this period (of order 160 years)
\citep{Dolukhanov:2005}.  We therefore argue that this timescale, of
order 600 years (and consequently also a lengthscale, of order 600
km), is the scale at which the spread of the Neolithic in Europe can
be considered well-modeled by a simple wave of advance: at longer
timescales and lengthscales (and clearly on the continental scale),
such a model of the spread performs well; at shorter timescales and
lengthscales, significant local deviations from such a simple spread
must be expected.  The quantification of this scale is an important
result from our inference.

Of course, the conclusions above must depend to some extent upon the
specific models introduced here (both our mathematical wavefront model
and the statistical model involving a global error parameter).  In
extensions to this work, we intend to investigate the robustness of
these results with respect to various changes in these models.
In our wavefront model, 
we first intend to investigate the possible importance of more regional
variations within the enhanced spread along waterways.
For example, we plan to allow different amplitudes of coastal enhancements 
within different regions,
thus allowing us to explore more effectively 
the possibility of a regionally enhanced spread in the Western Mediterranean,
as proposed for the Impressed Ware culture there.
We may also allow for advective velocities along other river systems,
in addition to the Danube and Rhine.

Implicit in our statistical model
is the assumption of spatially homogeneous normal errors. 
This assumption may be unnecessarily restrictive, 
and alternative statistical models (with more complex error structures)
should be considered.
For example, the statistical model may be adapted to allow 
for dates at different sites having differing uncertainties;
building this into the model might result in a more meaningful fit
(e.g.\ avoiding the possibility that a single global error parameter,
as used here, may be unnecessarily smoothing out the fit everywhere).
Further model refinements may also be possible.
The wave of advance clearly expects that nearby sites will have
similar arrival times;
in the current model, however, nearby sites are not linked in any way.
We will therefore allow for spatial correlation between nearby sites,
potentially helping to smooth out locally anomalous dates,
and also allowing another estimate of the scales over which 
the radiocarbon data correspond well to a simple wave of advance.

The Neolithisation of Europe is obviously not the only possible 
application for the methods introduced here, and applications to other regions
or to other prehistoric periods (e.g.\ the dispersal of palaeolithic cultures)
also have great potential.
Other applications would of course have their own difficulties,
with one likely challenge being the relative scarcity of empirical data
in many cases.  One such case is the spread of Neolithic culture from 
the Near East to South Asia;
there are significant gaps in the radiocarbon record between these regions,
and it would be of great interest to see how our methods could help 
to model this spread.

\bibliographystyle{pf}
\bibliography{my}

\appendix

\section{The Metropolis--Hastings algorithm}\label{app:met_hast}

We provide a detailed step-by-step description of the MCMC scheme we
use to sample from the posterior distribution of the model parameters,
$\pi(\boldsymbol{\theta},\sigma|\boldsymbol{t})$. 
(This type of scheme is well-established within the statistical 
literature \cite{gamerman2006markov,geman1984},
but is presented here to help the more general readership 
to appreciate the current work.)

We use a Gibbs sampling strategy, alternating between draws of the full
conditional distributions
$\pi(\sigma|\boldsymbol{\theta},\boldsymbol{t})$ and
$\pi(\boldsymbol{\theta}|\sigma,\boldsymbol{t})$. Algorithmically, we
perform the following steps:
\begin{enumerate}
\item Initialise $\sigma^{(0)}$ and $\boldsymbol{\theta}^{(0)}$. Set $j=1$.
\item Draw $\sigma^{(j)}\sim\pi(\,\cdot\,|\,\boldsymbol{\theta}^{(j-1)},\boldsymbol{t})$.
\item Draw $\boldsymbol{\theta}^{(j)}\sim\pi(\,\cdot\,|\,\sigma^{(j)},\boldsymbol{t})$.
\item Set $j:=j+1$ and go to step 2.
\end{enumerate}
The resulting Markov chain has invariant distribution given by 
$\pi(\boldsymbol{\theta},\sigma|\boldsymbol{t})$ \cite{gamerman2006markov}. 
The full conditional for $\sigma$ can be sampled straightforwardly as, if
$\zeta=\sigma^{-2}$ then
\begin{equation}
  \zeta|\boldsymbol{\theta},\boldsymbol{t}\sim \textrm{Gamma}\left( A, B \right),
\end{equation}
where 
\[
A=a+\frac{n}{2}, \quad
 B=b+\displaystyle \sum_{i=1}^{n} \left\{t_i-\tau(\mathbf{x}_i |\boldsymbol{\theta})\right\}^2/2.
\]
Hence, in step 2 of the Gibbs sampler, $\sigma^{(j)}$ is generated by
first drawing $\zeta^{(j)}|\boldsymbol{\theta}^{(j-1)},\boldsymbol{t}$
and then setting $\sigma^{(j)}=1/\sqrt{\zeta^{(j)}}$.  Since the full
conditional for $\boldsymbol{\theta}$ is analytically intractable we
use a Metropolis-Hastings update in step$~3$. Define
\[
\boldsymbol{\lambda}\equiv(\lambda_{1},\lambda_{2},\lambda_{3})^T =(\log(U_{0}),\log(V_{\rm C}),\log(V_{\rm R}))^T
\]
and note that under the prior specification adopted for
$\boldsymbol{\theta}$, each component $\lambda_{i}$, $i=1,2,3$ follows
a normal distribution (independently) \emph{a priori}.  In step 3 of
the Gibbs sampler we propose a new value
$\boldsymbol{\lambda}^{*}$ via a symmetric random walk with normal
innovations, that is
\[
\lambda_{i}^{*}=\lambda_{i}+\omega_{i}\,,\qquad \omega_i \sim N(0,\delta_{i}^2)\,,\qquad i=1,2,3
\]
where the $\delta_{i}$ are tuning parameters, the choice of which will
influence the mixing of the Markov chain.  Large values of $\delta_{i}$ 
will lead to small acceptance probabilities, and the
chain will rarely move; whereas small $\delta_{i}$ will lead to many
accepted proposed values, but slow exploration of the parameter space.
We accept the proposed value and take
$\boldsymbol{\lambda}^{(j)}=\boldsymbol{\lambda}^*$ with probability
$\alpha$, otherwise we take the current value
$\boldsymbol{\lambda}^{(j)}=\boldsymbol{\lambda}^{(j-1)}$.  The
acceptance probability is given by
\begin{equation} \label{eq:accept}
\alpha=\min \left\{ 1, \frac{\pi(\boldsymbol{\lambda}^*)\pi(\boldsymbol{t}|\boldsymbol{\lambda}^{*},\sigma^{(j)})}{\pi(\boldsymbol{\lambda}^{(j-1)})
\pi(\boldsymbol{t}|\boldsymbol{\lambda}^{(j-1)},\sigma^{(j)})}\right\}
\end{equation}
where $\pi(\boldsymbol{\lambda})$ denotes the prior density ascribed
to $\boldsymbol{\lambda}$ and
$\pi(\boldsymbol{t}|\boldsymbol{\lambda},\sigma)$ is given by
Eq.~(\ref{eq:likelihood1}) with
$\boldsymbol{\theta}=\exp(\boldsymbol{\lambda})$.  

\section{Emulation}\label{app:emulator}

The MCMC inference scheme typically requires many iterations, with
each iteration requiring a full simulation of the expanding Neolithic
front to evaluate the likelihood function.  As simulations of
the front are computationally expensive, we emulate the
model using Gaussian processes (GP) \cite{SantnerWN03}, that is,
stochastic approximations to the arrival times obtained from the
wavefront model.
These methods are widely used in the computer models literature; see,
for example, \cite{kennedy01} and references therein.  In brief, the
wavefront model is run for a set of training points;
the emulator then allows the interpolation of the model output 
between these points.  For pragmatic reasons, we build an individual
emulator for each radiocarbon site, rather than attempt to build a
complex time-space emulator.

Consider the arrival time $\tau(\mathbf{x}_i |\boldsymbol{\theta})$ at a single site $i$. 
For simplicity of notation,
we denote this arrival time by $\tau(\boldsymbol{\theta})$. 
Our emulator for the arrival time 
uses a Gaussian process with mean $m(\cdot)$ and covariance function $k(\cdot,\cdot)$, that is
\begin{equation}\label{eq:GPrelation}
\tau(\cdot) \sim {\rm GP}({m}(\cdot),k(\cdot,\cdot)).
\end{equation}
We choose a suitable form for the mean function, given
the approximate relationship expected between the arrival time at a site 
and the parameter values, which are all speeds. The simple relation $d=Ut$,
where $d$ represents distance, $t$ time and $U$ speed, gives $t \propto 1/U$, 
so we choose a mean function which reflects this:
\begin{equation}
m(\boldsymbol{\theta})=\alpha_0+\mathbf{\alpha}_1\frac{1}{U_0}+\mathbf{\alpha}_2\frac{1}{V_{\rm R}}+\mathbf{\alpha}_3\frac{1}{V_{\rm C}},
\end{equation}
where the coefficients $\alpha_{k}$ are determined using least squares fits.
These coefficients essentially account for the relative importance of 
diffusive, river and coastal spread,
given the complicated geography between the source of the spread 
and the particular site being emulated.
There are various possible choices for the form of covariance function. 
We use a stationary Gaussian covariance function
\begin{equation}\label{eq:covariance}
k(\boldsymbol{\theta},\boldsymbol{\theta}')=a\exp\left(-\displaystyle\sum_{j=1}^3 \frac{(\theta_j-\theta_j')^2}{r^2_j} \right),
\end{equation}
with hyperparameters $a$ and $r_j$ ($j=1,2,3$), 
which must be determined from the training data.

Suppose that $p$ simulations of the (computationally expensive)
wavefront model are available to us, each providing the arrival time
at each radiocarbon-dated site.  Let $\boldsymbol{\tau}(\Theta)
=(\tau(\btheta_{1}),\ldots,\tau(\btheta_{p}))^{T}$ denote the
$p$-vector of arrival times resulting from the wavefront model with
input values $\Theta=(\boldsymbol{\theta}_1,\ldots,
\boldsymbol{\theta}_{p})^{T}$, where
$\btheta_i=(U_{0,i},V_{{\rm R},i},V_{{\rm C},i})^T$. 
A Gaussian process can be viewed as an infinite collection of random variables, 
any finite number of which are jointly normally distributed. 
Therefore, from Eq.~(\ref{eq:GPrelation}), we have
\begin{equation*}\label{eq:GPrelation2}
\boldsymbol{\tau}(\Theta)\sim N(\boldsymbol{m}(\Theta),K({\Theta},{\Theta})),
\end{equation*}
where $\boldsymbol{m}(\Theta)$ is the mean vector with $j$th
element $m(\boldsymbol{\theta}_{j})$, and
$K({\Theta},{\Theta})$ is the variance matrix with $(j,\ell)$th
element $k(\boldsymbol{\theta}_{j},\boldsymbol{\theta}_\ell)$.

We can model the front arrival time at the site
for other values of the input parameters, $\boldsymbol{\theta}^*$, as follows. 
Using the standard properties of
the multivariate normal distribution, the arrival time has
distribution
\begin{equation}
\tau(\boldsymbol{\theta}^{*}) |\boldsymbol{\tau}(\Theta) \sim N\left({\mu}(\boldsymbol{\theta}^*),{\Sigma}(\boldsymbol{\theta}^{*}) \right),
\end{equation}
where
\begin{align*}
\mu(\boldsymbol{\theta}^*) 
&= m(\boldsymbol{\theta}^*)+K(\boldsymbol{\theta}^*,{\Theta})\left[K({\Theta},{\Theta})\right]^{-1}\left[ \boldsymbol{\tau}(\Theta)-\boldsymbol{m}(\Theta)\right]\\
\Sigma(\boldsymbol{\theta}^{*}) 
&= K(\boldsymbol{\theta}^*,\boldsymbol{\theta}^*)-K(\boldsymbol{\theta}^*,{\Theta})\left[K({\Theta},{\Theta})\right]^{-1}K(\boldsymbol{\Theta},\boldsymbol{\theta}^*).
\end{align*}
To simplify the notation, we have dropped the dependence in
these expressions on the hyperparameters $a$ and $r_j$ ($j=1,2,3$).

\subsection{Fitting the emulator}

We build a separate emulator for each site for which we have radiocarbon data.  
Although a single run of the wavefront model for particular input
parameters~$\boldsymbol{\theta}_i$ is computationally intensive, 
such a run gives the first arrival time at all sites,
so that only $p$ runs of the wavefront model are needed to
construct the training data for all $n$ emulators.

\begin{figure*}
\begin{center}
    \includegraphics[width=0.5\textwidth]{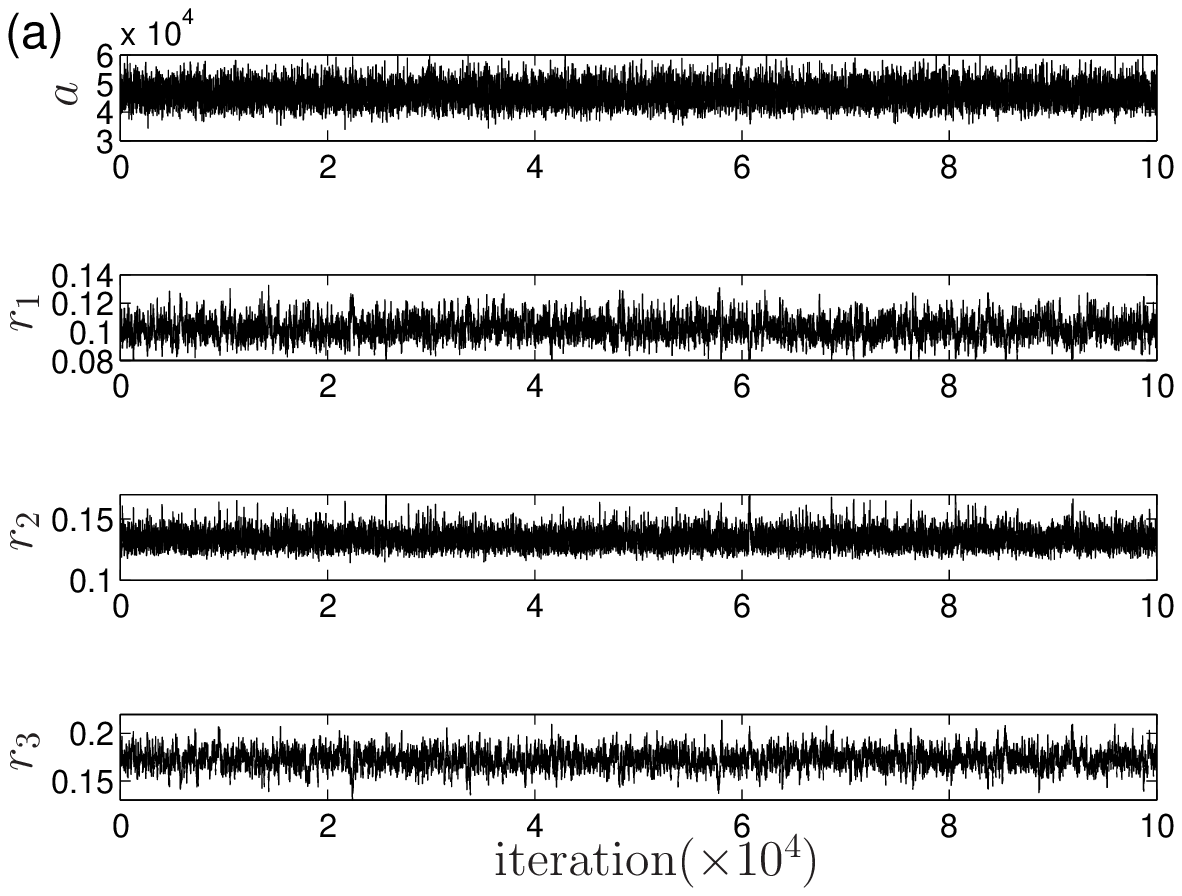}
\hfill
    \includegraphics[width=0.45\textwidth]{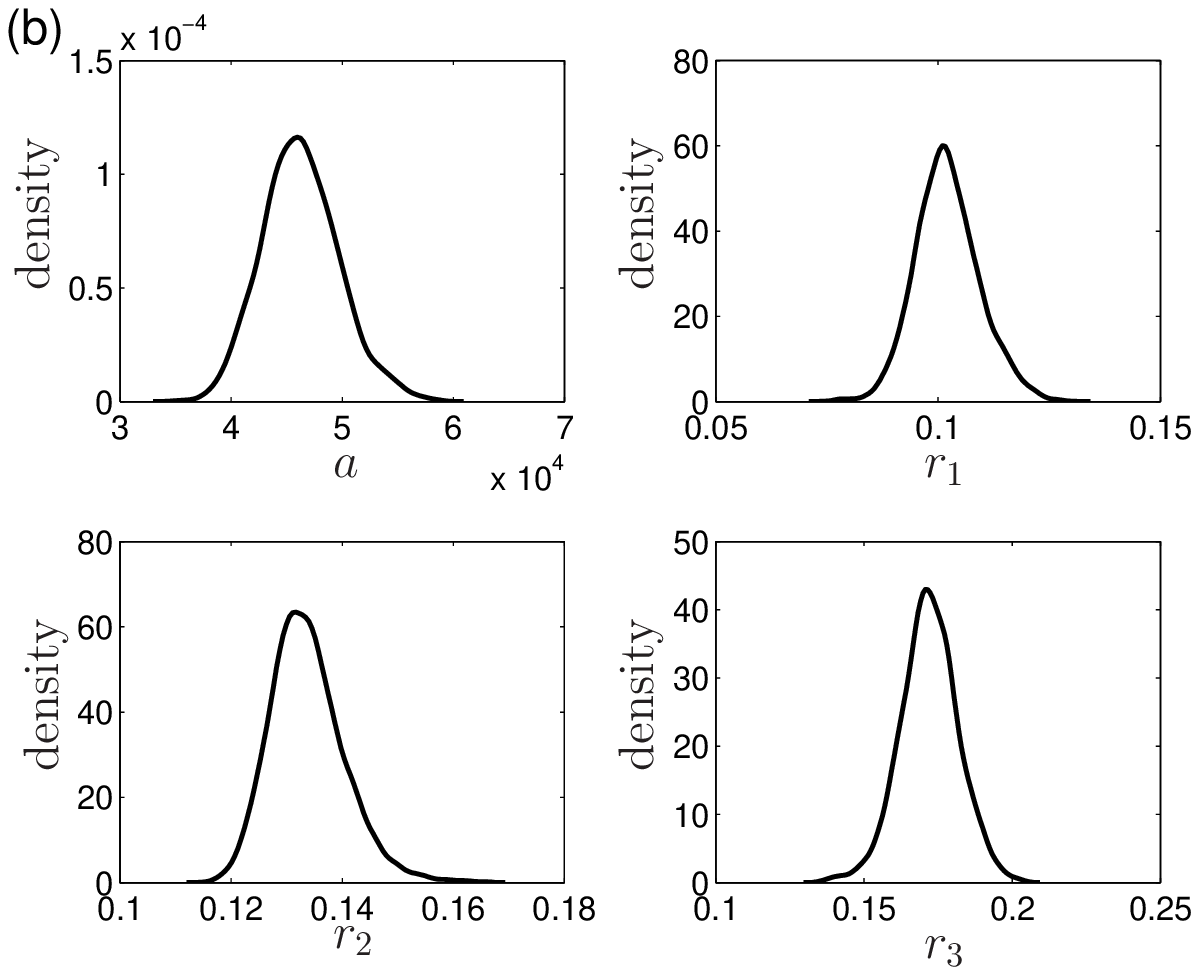}
    \caption{Traces for the hyper-parameters of the emulator for the radiocarbon site Achilleion, located at latitude 39.2N, longitude 22.38E (a),
and the posterior distributions of the four hyper-parameters 
from these MCMC chains (b).
The output from these chains varies from site to site, and it is important to construct a separate emulator for each site, to obtain accurate emulation of the wavefront model. }
\label{fig:hyperparam}
\end{center}
\end{figure*}

We fit each emulator using a Metropolis--Hasting algorithm (similar to
that described in Appendix~\ref{app:met_hast}), to obtain the posterior
distributions for the hyperparameters.  Fig.~\ref{fig:hyperparam}
(left) shows the traces of the resulting hyper-parameter chains (for a
single radiocarbon site), and Fig.~\ref{fig:hyperparam} (right) shows
the corresponding posterior densities.  These plots are representative
of the MCMC estimation of the posterior hyperparameter distributions
at other radiocarbon sites.

Whilst it is possible in theory to fit the emulator 
and the statistical model in Eq.~(\ref{eq:likelihood1}) jointly using
an MCMC scheme, this would be extremely computationally
expensive. We therefore fit the emulator and the statistical model
separately.  In particular, when using the emulator output in the
inference scheme described in Appendix~\ref{app:met_hast}, we fix the
hyperparameters at their posterior means.  We believe this approach is
justified, as even allowing for the (low) posterior uncertainty of the
hyperparameters makes little difference to the (predictive) fit of the
emulators.


\begin{figure}
\begin{center}
    \includegraphics[width=0.48\textwidth]{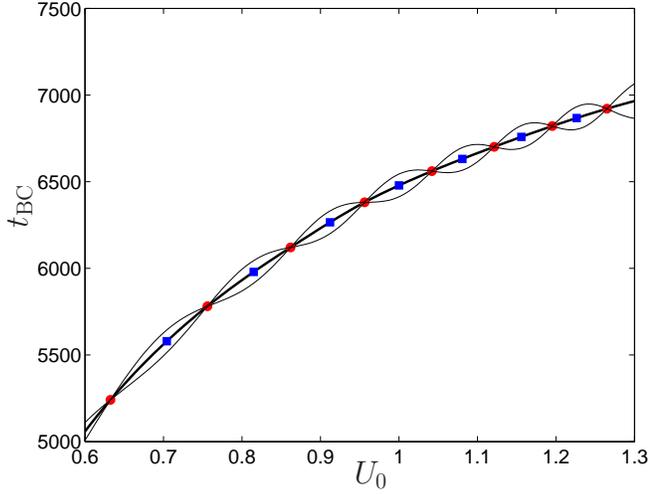}
    \caption{Color online) The mean output from one emulator (thick black line), with the training points used also plotted (red circles). The thin upper and lower lines show $\pm$ two standard deviations of the emulator output. The (blue) squares represent further output from the wavefront model, which were not used in constructing the emulator. The discrepancy between these points and the emulator output can be used as a test of the emulator.}
\label{fig:nu_emulation}
\end{center}
\end{figure}

For illustration, 
Fig.~\ref{fig:nu_emulation} shows the output from one emulator, 
together with the training data, 
when we fix two of the parameters ($V_{\rm C}$ and $V_{\rm R}$) 
and consider only variations in the $U_0$-axis.
The magnitudes of the errors shown in the plot 
are consistent with those from the three-parameter emulator.


\subsection{Selection of training points}

The selection of the training points ${\Theta}$ in parameter space
merits further comment.
Although using a regular lattice design is appealingly simple, it is
not particularly efficient. Instead, we adopt a more commonly used
design for fitting Gaussian processes, the Latin Hypercube
Design (LHD) \citep{McKay:1979}. Designs of this class distribute points 
within a hypercube in parameter space more efficiently than a lattice design.
If we consider any single parameter direction in isolation, 
the mean separation between points is $p^{-1}$, 
as opposed to $p^{-1/3}$ for a regular lattice.

We constructed our 200-point LHD using the \textsc{Matlab} routine
\texttt{lhsdesign}.  Initially we set the lower bounds of the hypercube to
be the origin and used the upper 1 percentiles of the prior
distribution as its upper bounds.  
We then repeatedly ran the inference algorithm 
(described in the following section), used the results to determine 
a conservative estimate of a hypercube containing all points in the MCMC output
(and therefore plausibly containing all of the posterior density), 
and generated another LHD.  The final LHD used for inferences on 
$(U_{0},V_{\rm C},V_{\rm R})^T$
in this paper is contained within the hypercube
$(0,3.1)\times(0,3)\times(0,2)$.

\subsection{Testing the emulator}

It is imperative that the accuracy of the fitted emulators as an 
approximation to the wavefront model be assessed. We therefore 
considered various quantitative statistics \cite{Bastos:2009}. 
We created a second LHD with $p^{*}=100$ points,
$\Theta^{*}=(\btheta^{*}_{1},\ldots ,\btheta^{*}_{p^{*}})^{T}$,
and determined the front arrival time
at all sites for each $\btheta^{*}_{i}$ using both the emulator mean
(with its hyperparameters fixed at their posterior means) and the
wavefront model: 
we denote these arrival times by
$\boldsymbol{\tau}^{*}$ and $\boldsymbol{\tau}$ respectively.
(Separate values of these quantities exist for all sites;
but for simplicity, as in the preceding sections, 
the specialization to individual sites is left implicit.)

\begin{figure}
\bigskip

\begin{center}
    \includegraphics[width=0.48\textwidth]{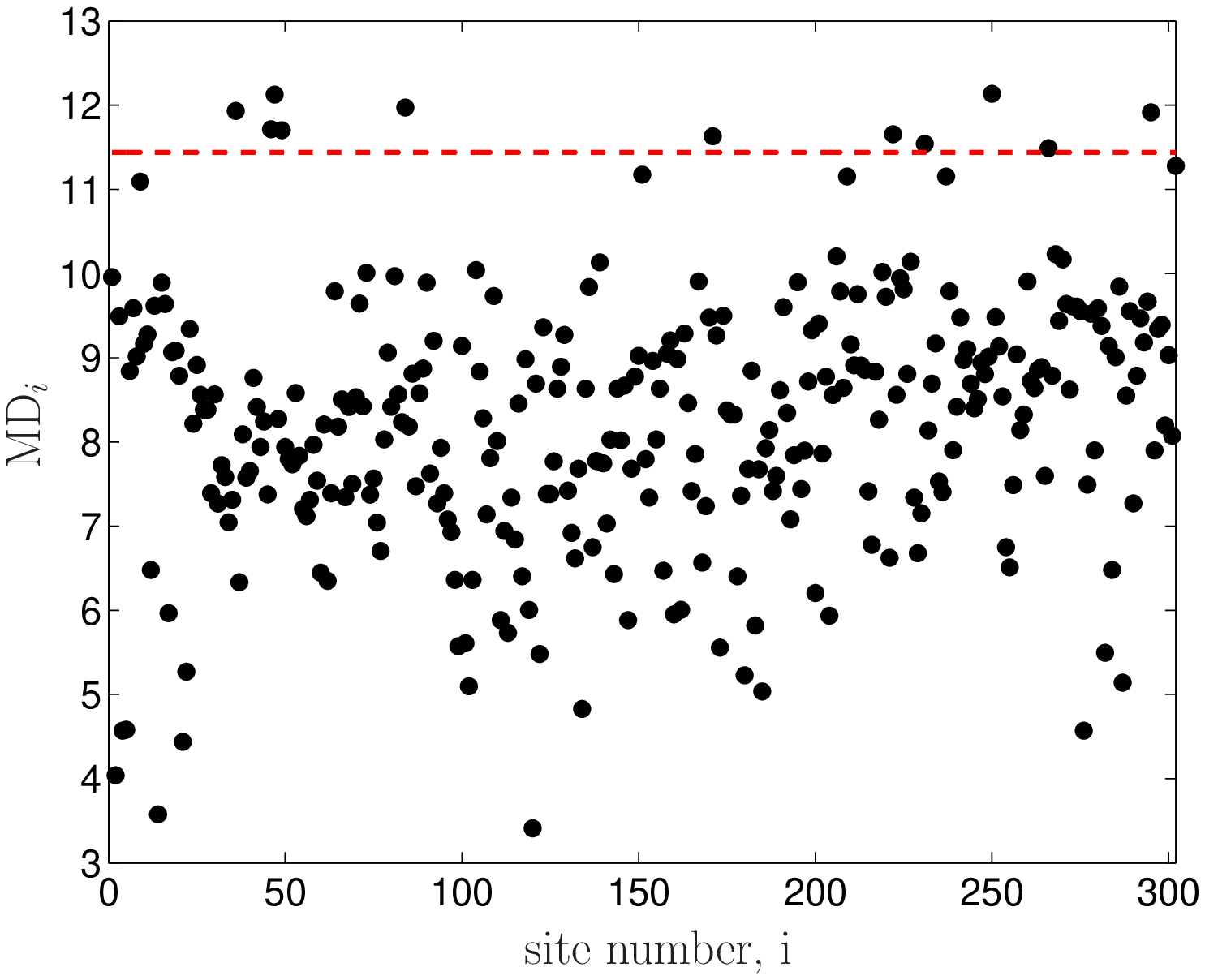}
    \caption{(Color online) Test of the fit of the emulators to the radiocarbon data: the Mahalanobis distance $\textrm{MD}_{i}$ is plotted for each radiocarbon site $i$. The horizontal (red) dashed line (at ${\rm MD}_i=11.44$)
marks the upper $95\%$ point of the Mahalanobis distance distribution.}
\label{fig:emulator_test}
\end{center}
\end{figure}

For brevity, 
we discuss only the analysis of a statistic which includes 
both site-specific accuracy
and correlation between residual errors (at the emulator test points):
the Mahalanobis distance,
$\textrm{MD}$, defined via
\begin{equation}
\textrm{MD}^2=(\boldsymbol{\tau}^* - \boldsymbol{\tau})^{T} V(\boldsymbol{\Theta}^{*})^{-1} (\boldsymbol{\tau}^* - \boldsymbol{\tau})\,,
\end{equation}
where
\begin{equation*}
V(\Theta^*) = K(\Theta^*,\Theta^*)-K(\Theta^*,\Theta)K(\Theta,\Theta)^{-1}K(\Theta,\Theta^*) .
\end{equation*}
(Note that $V(\Theta^*)$ is defined analogously to
$\Sigma(\boldsymbol{\theta}^{*})$ above, but now contains information
about all $p^*$ test points in $\Theta^*$.)  It can be shown that
MD$^2$ follows a scaled $F$-distribution \cite{Bastos:2009} in the
case of the GP emulator,
with $\textrm{MD}^2\sim p^*(p-5)F_{p^{*},\;p-3}/(p-3)$.
Figure~\ref{fig:emulator_test} shows the Mahalanobis distance at each
site, together with the upper $95\%$ point of its distribution.  This,
along with our analyses of other statistics (not presented here), confirms
that the emulators provide a reasonable fit throughout the design space. 
These diagnostics gave similar results for different LHDs,
without any systematic site-specific biases.

\end{document}